\documentclass[onecolumn,aps,superscriptaddress,showpacs]{revtex4}

\usepackage{amssymb}
\usepackage{amsmath}
\usepackage{graphicx}
\usepackage[normalem]{ulem}
\usepackage[dvips]{color}
\usepackage{verbatim}
\usepackage{rotating}

\begin{document}
\title{Understanding transport simulations of heavy-ion collisions at 100 and 400 AMeV\\
\small $-$ A comparison of heavy ion transport codes under controlled conditions}

\author{Jun Xu}
\email{xujun@sinap.ac.cn} \affiliation{Shanghai Institute of Applied
Physics, Chinese Academy of Sciences, Shanghai 201800, China}
\author{Lie-Wen Chen}
\email{lwchen@sjtu.edu.cn} \affiliation{Department of Physics and
Astronomy and Shanghai Key Laboratory for Particle Physics and
Cosmology, Shanghai Jiao Tong University, Shanghai 200240, China}
\author{ManYee Betty Tsang}
\email{tsang@nscl.msu.edu} \affiliation{National Superconducting
Cyclotron Laboratory and Department of Physics and Astronomy,
Michigan State University, East Lansing, Michigan 48824, USA}
\author{Hermann Wolter}
\email{hermann.wolter@physik.uni-muenchen.de}
\affiliation{Fakult\"at f\"ur Physik, Universit\"at M\"unchen, D-85748 Garching, Germany}
\author{Ying-Xun Zhang}
\email{zhyx@ciae.ac.cn} \affiliation{China Institute of Atomic
Energy, Beijing 102413, P.R. China}

\author{Joerg Aichelin}
\affiliation{SUBATECH, UMR 6457, Ecole des Mines de Nantes -
IN2P3/CNRS - Universit\'e de Nantes, France}
\author{Maria Colonna}
\affiliation{INFN-LNS, Laboratori Nazionali del Sud, 95123 Catania,
Italy}
\author{Dan Cozma}
\affiliation{IFIN-HH, Reactorului 30, 077125 M\v{a}gurele-Bucharest,
Romania}
\author{Pawel Danielewicz}
\affiliation{National Superconducting Cyclotron Laboratory and
Department of Physics and Astronomy, Michigan State University, East
Lansing, Michigan 48824, USA}
\author{Zhao-Qing Feng}
\affiliation{Institute of Modern Physics, Chinese Academy of
Sciences, Lanzhou 730000, People's Republic of China}
\author{Arnaud Le F\`{e}vre}
\affiliation{GSI Helmholtzzentrum f\"{u}r Schwerionenforschung, Darmstadt, Germany}
\author{Theodoros Gaitanos}
\affiliation{Department of Theoretical Physics, Aristotle University of Thessaloniki, GR-54124 Thessaloniki, Greece}
\author{Christoph Hartnack}
\affiliation{SUBATECH, UMR 6457, Ecole des Mines de Nantes -
IN2P3/CNRS - Universit\'e de Nantes, France}
\author{Kyungil Kim}
\affiliation{Rare Isotope Science Project, Institute for Basic
Science, Daejeon 305-811, Korea}
\author{Youngman Kim}
\affiliation{Rare Isotope Science Project, Institute for Basic
Science, Daejeon 305-811, Korea}
\author{Che-Ming Ko}
\affiliation{Cyclotron Institute and Department of Physics and
Astronomy, Texas A$\&$M University, College Station, Texas 77843,
USA}
\author{Bao-An Li}
\affiliation{Department of Physics and Astronomy, Texas A$\&$M
University-Commerce, Commerce, TX 75429-3011, USA}
\author{Qing-Feng Li}
\affiliation{School of Science, Huzhou University, Huzhou 313000,
China}
\author{Zhu-Xia Li}
\affiliation{China Institute of Atomic Energy, Beijing 102413, P.R.
China}
\author{Paolo Napolitani}
\affiliation{IPN, CNRS/IN2P3, Universit\'e Paris-Sud 11, 91406 Orsay
cedex, France}
\author{Akira Ono}
\affiliation{Department of Physics, Tohoku University, Sendai
980-8578, Japan}
\author{Massimo Papa}
\affiliation{INFN - Sezione di Catania, 95123
Catania, Italy}
\author{Taesoo Song}
\affiliation{Frankfurt Institut for Advanced Studies and Institute
for Theoretical Physics, Johann Wolfgang Goethe Universit\"at,
Frankfurt am Main, Germany}
\author{Jun Su}
\affiliation{Sino-French Institute of Nuclear Engineering $\&$
Technology, Sun Yat-sen University, Zhuhai 519082, China}
\author{Jun-Long Tian}
\affiliation{College of Physics and Electrical Engineering, Anyang
Normal University, Anyang, Henan 455000, China}
\author{Ning Wang}
\affiliation{Department of Physics and Technology, Guangxi Normal
University, Guilin 541004, China}
\author{Yong-Jia Wang}
\affiliation{School of Science, Huzhou University, Huzhou 313000,
China}
\author{Janus Weil}
\affiliation{Frankfurt Institut for Advanced Studies and Institute
for Theoretical Physics, Johann Wolfgang Goethe Universit\"at,
Frankfurt am Main, Germany}
\author{Wen-Jie Xie}
\affiliation{Department of Physics, Yuncheng University, Yuncheng
044000, China}
\author{Feng-Shou Zhang}
\affiliation{Key Laboratory of Beam Technology and Material
Modification of Ministry of Education, College of Nuclear Science
and Technology, Beijing Normal University, Beijing 100875, China}
\author{Guo-Qiang Zhang}
\affiliation{Shanghai Institute of Applied Physics, Chinese Academy
of Sciences, Shanghai 201800, China}

\begin{abstract}
Transport simulations are very valuable for extracting physics
information from heavy-ion collision experiments. With the
emergence of many different transport codes in recent years, it
becomes important to estimate their robustness in extracting physics
information from experiments. We report on the results of a
transport code comparison project. 18 commonly used transport codes
were included in this comparison: 9 Boltzmann-Uehling-Uhlenbeck-type
codes and 9 Quantum-Molecular-Dynamics-type codes. These codes have
been required to simulate Au+Au collisions using the same physics
input for mean fields and for in-medium nucleon-nucleon cross
sections, as well as the same initialization set-up, the impact
parameter, and other calculational parameters at 100 and 400 AMeV
incident energy. Among the codes we compare one-body observables
such as rapidity and transverse flow distributions. We also monitor
non-observables such as the initialization of the internal states of
   colliding nuclei and their stability, the collision rates and the Pauli blocking. We find that not completely identical initializations constitute partly for different evolutions. Different strategies to determine the collision probabilities, and to enforce the Pauli blocking, also produce considerably different results. There is a substantial spread in the predictions for the observables, which is much smaller at the higher incident energy. We quantify the uncertainties in the collective flow resulting from the simulation alone as about $30\%$ at 100 AMeV and $13\%$ at 400 AMeV, respectively. We propose further steps within the code comparison project to test the different aspects of transport simulations in a box calculation of infinite nuclear matter. This should, in particular, improve the robustness of transport model predictions at lower incident energies where abundant amounts of data are available.
\end{abstract}


\maketitle

\section{Introduction}
\label{intro}

Understanding the behavior of nuclear systems in a wide range of
densities, temperatures, and proton-neutron asymmetries,
characterized by the equation of state (EoS) of nuclear matter, is
one of the major goals in nuclear physics research. As the main
characteristic of bulk nuclear matter, the EoS is also an important
input to the study of astrophysical objects or processes like
neutron stars or core collapse supernovae. In the laboratory, the
EoS has been studied with heavy-ion collisions (HIC), which can
create a wide range of density and energy conditions depending on
the incident energy of the collision, the size of the colliding
system, and impact parameters. The EoS is not observed directly, but
has to be inferred from the properties of reaction products. At
lower energies, heavy-ion collisions are interpreted with
considerable success by statistical models. Since heavy-ion
collisions are dynamical processes, the preferred method relies on
non-equilibrium theories, which model the reaction and have as inputs
the EoS, effective nucleon-nucleon (NN) cross sections, effective
nucleon masses, and other physics quantities used in the models. For
the energies under consideration here, i.e., from the Fermi energy
regime to relativistic energies, transport theories are an important
tool to obtain valuable information from heavy-ion collisions.

Transport theories have been used for many years, starting from
 the early works of Refs.~\cite{Ger84,Kruse85,Aic86},  to interpret heavy-ion
collisions. Their use has achieved remarkable success. For example, the EoS has been constrained
rather well for symmetric nuclear matter~\cite{kaon,Dan02,Fuchs06,LeFevre16} with
transport models. On the other hand, the isovector sector of the
EoS, i.e., the nuclear symmetry energy or Asy-EoS, which is of great
importance for the astrophysical applications, is still less known. Even though constraints for the
symmetry energy are becoming increasingly
stringent~\cite{Bar05,Ste05,Lat07,BCK08,Tsa11,Lat12,Hor14}, there are
still large uncertainties particularly above saturation densities. For the investigation of the
symmetry energy which contributes only a fraction of the total
energy, but the influence of which increases with the asymmetry of
nuclear systems, an increased precision of the prediction of
transport theories is required (for a recent review, see Ref.~\cite{BALi14}).

Recently, different transport models have given different
predictions for physical observables with seemingly similar nuclear
input. Considering the different approaches to transport theory, it
is important to disentangle the causes that lead to different
predictions via, e.g., the transport code comparison project. One of the goals of the code comparison project is to
establish a theoretical systematic error that quantifies the model
dependence of transport predictions. Our eventual goal is to minimize this error even if it may not be possible
to get an exact convergence of the results from different transport
codes.

Possible reasons for the model dependence are inherent in the
complexity of transport calculations. Basically two families of
transport approaches are used in the study of heavy-ion collisions.
One is the Boltzmann-Vlasov type, which is formulated for the
evolution of the one-body phase-space density under the influence of
a mean field. The other is the molecular dynamics type, which is
formulated in terms of nucleon coordinates and momenta under the
action of a many-body Hamiltonian. Both are supplemented with a
two-body collision term. There are many variants of these two basic
approaches. We will refer to the first type of theories collectively
as Boltzmann-Uehling-Uhlenbeck (BUU) theories, and to the second as
Quantum Molecular Dynamics (QMD) theories, according to their most
common representatives. The equations of these theories are generally solved by numerical simulations due to their complexity. Different strategies are used in BUU- and QMD-type models, but also in
individual codes within each family. These differences are not
always evident in publications when results are presented.
Simulations to interpret experimental data often employ different
physical inputs and slightly different conditions, such that the
results may not be directly compared, and the model uncertainties
cannot be separated from the variations of the physical input. To
provide a better understanding of these differences, all
calculations shown in the present work use exactly the same physical
input and, as closely as possible, the same initial conditions.

It has long been recognized that a comparison between different
transport models is very much needed. In 2004 the community met at
the European Center for Theoretical Studies for Nuclear Physics and
Related Fields (ECT*) in Trento, Italy, to compare mainly the
particle production and also yields, rapidity distributions, and transverse spectra
 from various transport codes in the 1 AGeV
regime. Results from that comparison were published in
Ref.~\cite{Kol05}. Following this first step, an attempt was made to
compare bulk observables, namely anisotropies of momentum
distributions (flow) and collision rates, at lower energies (100 and
400 AMeV) in a workshop in Trento in 2009. While anisotropies of
momentum distributions are widely used in the analyses of HIC, it
became evident that their prediction is much less robust than that of the
particle production at higher energies. The Trento workshop was
followed by the International Workshop on Simulations of Low- and
Intermediate-Energy Heavy-Ion Collisions in 2014, in Shanghai,
China~\cite{transport2014}. Before, during, and after the meeting,
we have compared the calculated results based on 18 transport codes
listed in Table~\ref{T1}~\cite{GXNU-IQMD}. This list includes the
transport codes most widely used by the intermediate-energy
heavy-ion community today.

The comparison project, including homework results as well as details of the different codes, will be published by Springer in a special Volume. This
article intends to present in a timely manner the relevant results
obtained in the first phase of the comparison project, before the
book becomes available. Response to this article will be used to
improve the contents of the book. Furthermore, we hope that this
project will lead to a useful milestone making it possible to
better clarify different strategies adopted in the formulation of
the transport models.

This article is organized as follows: Brief descriptions of the two
families of the most widely used transport models, BUU and QMD, will
be given in Sec.~\ref{theory}. Section~\ref{homework} gives details
of the set-up of the calculations, called here the homework of the
code comparison, since we asked each participant to provide results
of calculations with specified requirements. In
Sec.~\ref{initialization} the initialization of the collisions will
be discussed together with the stability of the initial set-up. This
is realized by performing a calculation with a large impact
parameter, $\text{b}=20$ fm, such that an actual nucleus-nucleus
collision rarely occurs. In Sec.~\ref{results} we compare the
results for a more realistic heavy-ion collision at an intermediate
impact parameter, $\text{b}=7$ fm, to study different
observables and collision rates. Here we compare collisions at two
energies, 100 and 400 AMeV. At different energies the influence of
the mean field and collisions will be different. In fact, it will be
seen, that collisions at 100 AMeV represent a particularly critical
regime, where there is a strong competition between the mean field
and the collisions, and where, therefore, the difference in the
codes is magnified. This appears less critical at the higher energy.
In Sec.~\ref{discussion} a critical discussion of the aims and
implications of the present code comparison is presented. Finally in
Sec.~\ref{summary} we summarize the results and achievements of this
investigation. We also discuss further steps to reach the goal of a
better convergence and understanding of different transport models,
for which we plan a follow-up calculation for an infinite system of
nuclear matter, set up as a calculation in a box with periodic
boundary conditions.

In this paper we show some highlight results of the code comparison
project, while technical details of all the codes used in the
comparison will appear in the book mentioned above. The
relevant ingredients of the codes used in the present code comparison project are
summarized in Tables~\ref{apt1} and \ref{apt2} containing the
information for initialization, nucleon-nucleon scatterings and
Pauli Blocking, where appropriate. There are considerable
differences in calculating occupation probabilities and treating the
Pauli blocking. We stress that the treatment given in these tables
is not necessarily the mode in which they are used in applications
to data of heavy-ion collisions, which will be given in the book. Until the special volume on the "Transport Code Comparison
Project" is published, detailed descriptions of all the codes can be
found in the references provided in the tables.

\begin{table}[h]\small
  \centering
  \caption{The names, authors and correspondents, and representative references of 9 BUU-type and 9 QMD-type models participating in the transport code comparison project. The intended beam-energy range for each code is given in GeV.
}
    \begin{tabular}{c|c|c|c|c|c|c|c|c}
    \hline
    BUU-type & code correspondents & energy range & reference  &&  QMD-type & code correspondents & energy range & reference\\
    \hline
     BLOB    & P.Napolitani,M.Colonna & 0.01 $\sim$ 0.5 & \cite{BLOB} &&AMD & A.Ono & 0.01 $\sim$ 0.3 & \cite{AMD} \\
     GIBUU-RMF & J.Weil & 0.05 $\sim 40$ & \cite{GIBUU}  && IQMD-BNU  & J.Su,F.S.Zhang & 0.05 $\sim 2$ & \cite{IQMD-BNU} \\
     GIBUU-Skyrme & J.Weil & 0.05 $\sim$ 40 & \cite{GIBUU}   && IQMD & C.Hartnack,J.Aichelin  & 0.05 $\sim$ 2 & \cite{Har89,Aic91}  \\
     IBL    &  W.J.Xie,F.S.Zhang & 0.05 $\sim$ 2 & \cite{IBL} && CoMD  & M.Papa & 0.01 $\sim$ 0.3 & \cite{CoMD}\\
     IBUU   &  J.Xu,L.W.Chen,B.A.Li & 0.05 $\sim$ 2 & \cite{BCK08,IBUU} && ImQMD-CIAE & Y.X.Zhang,Z.X.Li & 0.02 $\sim$ 0.4 & \cite{ImQMD-CIAE} \\
     pBUU   &  P.Danielewicz & 0.01 $\sim$ 12 & \cite{pBUU} && IQMD-IMP & Z.Q.Feng & 0.01 $\sim$ 10 & \cite{IQMD-IMP}\\
     RBUU   & K. Kim,Y.Kim,T.Gaitanos& 0.05 $\sim$ 2 & \cite{RBUU} && IQMD-SINAP & G.Q.Zhang & 0.05 $\sim$ 2 & \cite{IQMD-SINAP} \\
     RVUU   & T.Song,G.Q.Li,C.M.Ko & 0.05 $\sim$ 2 & \cite{RVUU} && TuQMD & D.Cozma & 0.1 $\sim$ 2 & \cite{TuQMD}\\
     SMF    & M.Colonna,P.Napolitani & 0.01 $\sim$ 0.5 & \cite{SMF} && UrQMD & Y.J.Wang,Q.F.Li & 0.05 $\sim$ 200 & \cite{UrQMD,Bas98}\\
    \hline
    \end{tabular}
  \label{T1}
\end{table}

\begin{table}[h]
\newcommand{\tabincell}[2]{\begin{tabular}{@{}#1@{}}#2\end{tabular}}
\caption{Initialization and nucleon-nucleon scattering treatment used in various codes in homework calculations.}
\begin{tabular}{l|l|l|l|l|l|l}
\hline
\textbf{Code name} & \tabincell{c}{Shape of \\ particles} & $(\Delta x)^2$ [fm$^2$]\footnote{$\Delta x$ is the width of the Gaussian wavepacket as in Eq.~(\ref{phi}).} & \tabincell{c} {$\delta<r^2>^{1/2}$ (fm)}\footnote{$\delta<r^2>^{1/2}=<r^2>^{1/2}-<r^2>^{1/2}_{WS}$ with $<r^2>^{1/2}_{WS}$ from the required Woods-Saxon distribution.} & \tabincell{c} {$\delta<r^4>^{1/4}$ (fm)}\footnote{$\delta<r^4>^{1/4}=<r^4>^{1/4}-<r^4>^{1/4}_{WS}$ with $<r^4>^{1/4}_{WS}$ from the required Woods-Saxon distribution.} & Attempted collisions & \tabincell{c}{1st collisions \\ within same nucleus}  \\
\hline
\textbf{AMD} & Gaussian &1.56 & -0.01 & 0.01  & $p=\alpha e^{-\nu R_{ij}^2}v_{ij}\Delta t$ & yes \\
\hline
\textbf{IQMD-BNU } & Gaussian & 1.97 & 0.32 & 0.39 & Bertsch approach\footnote{"Bertsch approach" means: $b<\sqrt{\sigma^{med}/\pi}$ and $v_{ij}\gamma \Delta t/2 > |r_{ij} \cdot \vec{p}/p|$ as described in the Appendix B of Ref.~\cite{Ber88}.}&  no \\
\hline
\textbf{IQMD} & Gaussian &  2.16 & 0.64 & 0.85 & Bertsch approach &  yes\\
\hline
\textbf{CoMD} & Gaussian &  1.32 & -0.11 & -0.04 & $p=1-e^{\Delta t/\tau}$ &  yes \\
\hline
\textbf{ImQMD-CIAE} & Gaussian & 2.02 & 0.39 & 0.47 & Bertsch approach & yes \\
\hline
\textbf{IQMD-IMP} & Gaussian & 1.92 & 0.61 & 0.80 & Bertsch approach & yes \\
\hline
\textbf{IQMD-SINAP} & Gaussian & 2.16 & 0.03 & 0.12 & Bertsch approach & yes \\
\hline
\textbf{TuQMD} & Gaussian & $2.16$ & -0.17 & -0.17 & Bertsch approach & yes \\
\hline
\textbf{UrQMD} & Gaussian & 2 & 0.12 & 0.18 & collision time table\footnote{Details about the collision criterion in UrQMD can be found in Ref.~\cite{Bas98}} & yes \\
\hline
& \tabincell{c}{Shape of \\ test particle} & \tabincell{c}{$(\Delta x)^2$ [fm$^2$] \\or $l$ [fm]\footnote{$l$ is the lattice spacing for test particle with triangular shape. See its definition in Ref.~\cite{TF2}.}} & & & \\
\hline
\textbf{BLOB} & triangle & 2 & 0.10 & 0.07 & $p=\sigma^{med}\frac{(\rho_i+\rho_j)}{2} v_{ij}\Delta t$  & yes \\
\hline
\textbf{GIBUU-RMF} & Gaussian & 1 & -0.18 & -0.26 & Bertsch approach & yes \\
\hline
\textbf{GIBUU-Skyrme}  & Gaussian & 1 & -0.03 & -0.03 & Bertsch approach & yes \\
\hline
\textbf{IBL} & Gaussian & 2 & -0.32 & -0.42 & Bertsch approach & no \\
\hline
\textbf{IBUU}  & triangle & 1 & 0.01 & 0.04 & Bertsch approach & yes \\
\hline
\textbf{pBUU} & point & 0\footnote{The node separation for the calculation of average quantities is typically 0.92 fm, but can decrease with increasing energy. See Ref.~\cite{pBUU} for details.} & 0.01 & -0.02 & cell\footnote{See Ref.~\cite{Dan91} for details.} & yes  \\
\hline
\textbf{RBUU}  & invar.Gauss & 1.4 & -0.12 & -0.19 & Bertsch approach  & yes \\
\hline
\textbf{RVUU}  & point & 0 & 0.01 & 0.03 & Bertsch approach & yes \\
\hline
\textbf{SMF} & triangle & 2 & -0.13 & -0.18 & $p=\sigma^{med}\frac{(\rho_i+\rho_j)}{2} v_{ij}\Delta t$  & yes \\
\hline
\end{tabular}
\label{apt1}
\end{table}

\clearpage

\begin{table}[h]
\newcommand{\tabincell}[2]{\begin{tabular}{@{}#1@{}}#2\end{tabular}}
\caption{Pauli-blocking treatment used in various codes in homework calculations.}
\begin{tabular}{l|l|l|l}
\hline
\textbf{Code name} & \tabincell{c}{ Occupation probability $f_i$ }& \tabincell{c}{Blocking probability}\footnote{Occupation probability $f_i$ is replaced by min$[f_i,1]$ if $f_i$ is larger than 1.} & \tabincell{c}{Additional \\ constraints} \\
\hline
\textbf{AMD} & antisymmetrized wavepackets\footnote{See Ref.~\cite{AMD} for details.} &\tabincell{c}{physical wavepacket$^b$} & no\\
\hline
\textbf{IQMD-BNU} & \tabincell{c}{$f_i$ in $h^3$ } & \tabincell{c}{$1-(1-f_i)(1-f_j)$} & yes\footnote{Phase-space constraint, see Ref.~\cite{IQMD-BNU} for details.}\\
\hline
\textbf{IQMD}  & \tabincell{c}{$f_i$ in $h^3$ } & \tabincell{c}{$1-(1-f_i)(1-f_j)$} & yes\footnote{Isospin average, see Ref.~\cite{Har89} for details.}  \\
\hline
\textbf{CoMD}  & $f_i$ in $h^3$ & $f_i^{\prime}$, $f_j^{\prime}<f_{\mathrm{max}}=1.05-1.1$   & \tabincell{c}{yes\footnote{Phase-space constraint, see Ref.~\cite{Pap05} for details.}} \\
\hline
\textbf{ImQMD-CIAE} & \tabincell{c}{ $f_i$ in $h^3$ } & \tabincell{c}{$1-(1-f_i)(1-f_j)$}  & no \\
\hline
\textbf{IQMD-IMP}  & \tabincell{c}{$f_i$ in phase-space cell with \\$dx=3.367$ fm, $dp=89.3$ MeV/c} & \tabincell{c}{$1-(1-f_i)(1-f_j)$} & no \\
\hline
\textbf{IQMD-SINAP} & \tabincell{c}{${f_i} = \sum\limits_k {e^{ - {{({{\vec r}_k} - {{\vec r}_i})}^2}/\left[ {2(\Delta x)^2} \right]}}{e^{ - {{\left( {{{\vec p}_k} - {{\vec p}_i}} \right)}^2} \cdot 2 (\Delta x)^2/{\hbar ^2}}}$} & \tabincell{c}{$1-(1-f_i)(1-f_j)$} & no\\
\hline
\textbf{TuQMD} & \tabincell{c}{$f_i$ in spherical phase-space cell with \\$dx=3.0$ fm, $dp=240$ MeV/c\footnote{In TuQMD the Pauli Blocking is implemented by computing the
wave function overlap using the method described in Ref.~\cite{Aic91}.} } & $1-(1-f_i)(1-f_j)$ & \tabincell{c} {yes\footnote{Surface modification, see Ref.~\cite{Ach86} for details.}}  \\
\hline
\textbf{UrQMD} & \tabincell{c}{${f_i} = \sum\limits_k {e^{ - {{({{\vec r}_k} - {{\vec r}_i})}^2}/\left[ {2(\Delta x)^2} \right]}}{e^{ - {{\left( {{{\vec p}_k} - {{\vec p}_i}} \right)}^2} \cdot 2(\Delta x)^2/{\hbar ^2}}}$} & \tabincell{c}{$1-(1-f_i)(1-f_j)$}  & yes\footnote{Phase-space constraint: ${{4\pi } \over 3}r_{ik}^3{{4\pi } \over 3}p_{ik}^3 \ge  {\left( {{h \over 2}} \right)^3}/4$.}\\
\hline
& & & \\
\hline
\textbf{BLOB} & \tabincell{c}{ $f_i$ in sphere with radius 3.5 fm \\
with Gaussian weight in momentum space\footnote{Width of the
Gaussian from definition of test-particle agglomerates, see
Ref.~\cite{BLOB} for details.}} & \tabincell{c}{$1-(1-f_i)(1-f_j)$}
& yes\footnote{Wavepacket modulation (shape, widths) to ensure
strict
Pauli blocking.} \\
\hline
\textbf{GIBUU-RMF} & \tabincell{c}{ $f_i$ in phase-space cell with \\$dx=1.4$ fm, $dp=68$ MeV/c} & \tabincell{c}{$1-(1-f_i)(1-f_j)$} & no \\
\hline
\textbf{GIBUU-Skyrme} & \tabincell{c}{ $f_i$ in phase-space cell with \\$dx=1.4$ fm, $dp=68$ MeV/c} & \tabincell{c}{$1-(1-f_i)(1-f_j)$} & no \\
\hline
\textbf{IBL} & \tabincell{c}{$f_i$ in $h^3$ } & \tabincell{c}{$1-(1-f_i)(1-f_j)$} & yes\footnote{Fermi constraint, see Ref.~\cite{IBL} for details.}\\
\hline
\textbf{IBUU} & \tabincell{c}{ $f_i$ in phase-space cell with \\$dx=2.73$ fm, $dp=187$ MeV/c} & \tabincell{c}{$1-(1-f_i)(1-f_j)$ } & no \\
\hline
\textbf{pBUU} & $f_i$ in same and
neighboring spatial cell\footnote{See Ref.~\cite{pBUU} for details.}
& \tabincell{c}{$1-(1-f_i)(1-f_j)$}
& no \\
\hline
\textbf{RBUU} &  \tabincell{c}{ $f_i$ in phase-space cell with \\$dx=1.4$ fm, $dp=64$ MeV/c } & \tabincell{c}{ $ 1- (1-f_i)(1-f_j)$ }  & no\\
\hline
\textbf{RVUU} & \tabincell{c}{$f_i$ in phase-space cell with \\$dx=1.14$ fm, $dp=331$ MeV/c\footnote{Obtained using $dx=[3/(4\pi\rho_0)]^{1/3}$ and $dp=[6\pi^2\rho_0/(2s+1)]^{1/3}$, see Ref.~\cite{RVUU} for details.}} & \tabincell{c}{ $1-(1-f_i)(1-f_j)$ } & no\\
\hline
\textbf{SMF} & \tabincell{c}{ $f_i$ in sphere with radius 2.53 fm \\
with Gaussian weight in momentum space\footnote{The width of the Gaussian is 29 MeV/c.}} & \tabincell{c}{$1-(1-f_i)(1-f_j)$}  & no \\
\hline
\end{tabular}
\label{apt2}
\end{table}

\section{Brief description of transport theories}
\label{theory}

Transport theories are very useful for extracting physical information on
nuclear matter from heavy-ion collisions. In this section we briefly
characterize the two main approaches for transport simulations. This is
thus not intended as a comprehensive theoretical discussion of the
derivation and validity of transport theories, but rather as a
"technical" guide of the main characteristics, the methods of
implementation, and the ingredients and methods to perform transport
simulations in order to facilitate the discussion of the code
comparisons in this paper.

Transport theories describe the evolution of the one-body phase-space distribution in a heavy-ion collision under the action of a
mean field, two-body collisions, and their Pauli blocking. To which
extent higher-order correlations are taken into account will be
discussed below. There are essentially two approaches to solve the
problem: those, which evolve directly from the phase-space density,
generally called here Boltzmann-Uehling Uhlenbeck (BUU)
approaches~\cite{Ber88}, and those, which formulate the transport in
terms of nucleon coordinates and momenta, generally called here
Quantum Molecular Dynamics (QMD) approaches~\cite{Har89,Aic91}.
There are both relativistic and non-relativistic formulations of
each approach. Table~\ref{T1} lists the names, the authors and
correspondents, the intended energy range, and the main reference of
each code, that participates in this comparison.

\subsection{The Boltzmann-Uehling-Uhlenbeck approach}

The BUU theory can be derived from the
Born-Bogliubov-Green-Kirkwood-Yvon (BBGKY) hierarchy or from
some perspective more effectively, using the real time Green
function Martin-Schwinger formalism to arrive at the Kadanoff-Baym
equations~\cite{Malfliet90,Danielew84,Mao94}. With the
quasi-particle approximation (on-shell particles), an approximation
for the (complex) self energy, and a semi-classical approximation,
one arrives at the Boltzmann-Uehling-Uhlenbeck (BUU) equation for
the phase-space density. The equation describes the time evolution
of the one-body phase-space density $f(\vec{r},\vec{p};t)$
\begin{equation}\label{buueq}
\left( \frac{\partial}{\partial t} + \frac{ \vec{p}}{m} \cdot \nabla_r  - \nabla_r U \cdot \nabla_p \right) f(\vec{r},\vec{p};t) = I_{coll}[f;\sigma_{12}]
\end{equation}
under the influence of the mean field $U[f]$ (usually formulated as a
density functional depending on the baryon density as well as the isospin and possibly spin and assumed here
to be momentum independent) and a two-body collision term
\begin{eqnarray}\label{NNsca}
I_{coll} &=&  \frac{1}{(2\pi)^6}\int dp_2 dp_3 d\Omega |v-v_2| \frac{d\sigma_{12}^{med}}{d\Omega} (2\pi)^3 \delta(p+p_2-p_3-p_4) \notag \\
&\times&[f_3 f_4 (1-f) (1-f_2) - f f_2 (1-f_3) (1-f_4)]
\end{eqnarray}
where $\sigma^{med}$ is the in-medium nucleon-nucleon cross
section and assumed here to be elastic. The left-hand side of Eq.~(\ref{buueq}) is the Vlasov equation, which
can be derived as a semi-classical approximation to the time-dependent Hartree-Fock (TDHF) equation. The UU (Uehling-Uhlenbeck) abbreviation in the BUU equation stands for the introduction
of the Pauli blocking factors in the gain and loss terms of the
collision. Other names for the same equation are Vlasov-Uehling-Uhlenbeck (VUU),
Boltzmann-Nordheim-Vlasov (BNV), or Landau-Vlasov (LV).
The physical ingredients in the BUU equation are the mean field $U$ and
the in-medium cross section $\sigma^{med}$. The derivation
sketched above can give a consistent approximation for both
quantities, e.g., from the closed-time path Green function method as in Ref.~\cite{Mao94}. However, in many applications, and
also in this code comparison, these are specified independently.
This allows us to test which observables are sensitive to which
ingredients.

The BUU equation is a non-linear integral-differential equation
which cannot be solved analytically or in a direct numerical way.
Rather the common method is to simulate the solution by using the
test-particle (TP) technique, which was introduced to nuclear physics in the beginning of 80s by Wong~\cite{Won82} for the solution of the TDHF equation. Here the
(continuous) distribution function is resolved in terms of a (large)
number of discrete TPs as
\begin{equation}
f(\vec{r},\vec{p};t) = \frac{1}{N_{TP}}\sum_{i=1}^{N_{TP}A} g(\vec{r}-\vec{r}_i(t))\tilde{g}(\vec{p}-\vec{p}_i(t))
\end{equation}
where $A$ is the number of nucleons, $N_{TP}$ is the test particle
number per nucleon (100 in this work), $\vec{r}_i$ and $\vec{p}_i$ are the
coordinates and momenta of the test particles, and
$g$ and $\tilde{g}$ are the coordinate and momentum shapes. The coordinate shape is often taken as point particles ($\delta$
functions), but to reduce the number of test particles and make the
distribution smoother, it may be taken as Gaussians or
triangular shapes in some codes (see Table~\ref{apt1}). A finite shape in momentum space is not commonly used, but becomes relevant for momentum-dependent interactions or in the calculation of momentum dependent quantities. It can be shown, that approximatively under the
influence of the mean field, the test particle coordinates and momenta obey the
Hamiltonian equations of motion
\begin{equation}\label{canonical}
d\vec{r}_i/dt = \nabla_{\vec{p}_i} H; ~~~~~d\vec{p}_i/dt = -\nabla_{\vec{r}_i} H.
\end{equation}
One usually has to average
the density over a cell (typically of size 1 fm$^3$), which is
equivalent to a coarse-graining in the solutions of the equations.

The collision term is commonly simulated stochastically, performing
TP collisions with cross section $\sigma^\prime =
\sigma^{med}/N_{TP}$. At each time step the TP configuration is
sampled to find TPs which are closer than the geometrical limit
$d=\sqrt{\sigma^\prime/\pi}$ to make a collision. Sometimes
additional constraints are introduced, such as not allowing two
collisions of the same TP in the same time step, making sure that
the two TPs move towards each other, or requiring that the first
collision of two TPs are not in the same nucleus, etc (see
Table~\ref{apt2}). For each such "attempted" collision, the Pauli
blocking is checked by calculating the phase-space occupation for
the final states. Here an average over a final cell has to be taken
to obtain reasonably smooth results, also a type of coarse-graining.
The Pauli blocking probability is calculated in most cases from
$1-(1-f_3)(1-f_4)$ (see Table~\ref{apt2}). This still leads to an
over-occupation of a final cell. Some codes then disallow the
collision (or force the occupation to have the value 1). In
principle with Fermi statistics implemented at the beginning of the
reaction and the Pauli principle enforced in the collision term, the
Fermion nature of the system should be preserved in the evolution.
However, it has been shown~\cite{AbeAyik96} that the coarse-graining
can act as a dissipation which evolves the statistics to a classical
Maxwell-Boltzmann distribution.

The above method of solving the collision term, called the full
ensemble method, is numerically expensive since it scales like
$(AN_{TP})^2$. In most calculations the parallel ensemble
method is used, where the total number of test particles
is divided into $N_{TP}$ ensembles of $A$ particles each. Then collisions are allowed to occur only
   in each ensemble with cross section $\sigma^{med}$, while the Pauli
   blocking and the mean field are calculated by using the test
   particles from all the ensembles. It has been checked in
typical cases that this procedure gave results similar to the full
ensemble method.

\subsection{The Quantum Molecular Dynamics Approach}

The QMD approach to transport theory can be discussed from two
perspectives. It has a relation to classical molecular dynamics for
the nucleons with a Hamiltonian, which is formulated in terms of
two- or many-body interactions (but often also in terms of density
functionals). The nucleons are specified not as point particles but
as particles with finite widths usually of Gaussian shape. On the
other hand QMD can be derived from a time-dependent Hartree (TDH) theory with a trial wave function of a product of Gaussian single-particle wave functions $\phi_i(\vec{r};t)$ with positions
$\vec{R}_i(t)$ and momenta $\vec{P}_i(t)$ as variational parameters
\cite{Aic91}
\begin{eqnarray}\label{wf}
\Psi(\vec{r}_1, ..., \vec{r}_N;t) &=& \Pi \phi_i(\vec{r}_i;t), \\
\phi_i(\vec{r}_i;t) &=& \frac{1}{(2\pi)^{3/4} (\Delta x)^{3/2}} \exp\left[ -\frac{(\vec{r}_i-\vec{R}_i(t))^2}{(2\Delta x)^2} + i\vec{r}_i \cdot \vec{P}_i(t)\right].\label{phi}
\end{eqnarray}
The variation of the wave function, Eq.~(\ref{wf}), leads to
equations of motion for the nucleons which are very similar to BUU,
i.e., the same Eqs.~(\ref{canonical}) are valid for the centroids of
Gaussian wave function in coordinate and momentum space
$\vec{R}_i(t)$ and $\vec{P}_i(t)$, with $\vec{R}_i=\langle \vec{r}_i
\rangle$ and $\vec{P}_i=\langle \vec{p}_i \rangle$. Thus large differences
in the propagation are not expected for one-body observables. For
AMD (antisymmetrized molecular dynamics)~\cite{AMD} or FMD
(fermionic molecular dynamics)~\cite{FMD} approach, the
antisymmetrization of the wavepackets is taken into account, i.e.,
as derived from TDHF with a Slater determinant of Gaussians as a
trial wave function. The equations of motion become more
complicated, involving a norm matrix, since wavepacket overlap
changes when they move (for more details see Ref.~\cite{AMD}). A
variant of this approach is CoMD (constrained molecular dynamics),
which does not explicitly implement antisymmetrization, but takes
the effects into account in an effective interaction~\cite{CoMD}.
The single particle Gaussians in QMD usually have a fixed width,
while the change of the wavepacket shape is taken into account in a
version of AMD as the wavepacket splitting~\cite{Ono02}.

In QMD approaches the two-body collisions are also introduced, and
are simulated in much the same way as in the full ensemble technique
of BUU, see the preceding subsection that applies to QMD for the
nucleon coordinates and the full in-medium cross section
$\sigma^{med}$. In both cases nucleons (not test particles) collide
with the NN in-medium cross section. Thus a collision will affect
the distribution function considerably more than a TP collision in
BUU. The treatment of collisions in QMD approaches is intrinsically
stochastic. In contrast to BUU, two-nucleon collisions
   induce event-by-event fluctuations. A
large number of runs are performed (with different initial states),
which are considered as "events" and which are averaged to obtain the mean and the variation of the final result.

The above discussion also touches the question in which way any
higher-order correlations not captured by the one-body distribution
function are included in QMD approaches. Due to the form of the
trial wave function, QMD can probably describe classical N-body
correlations better than the BUU approach. On the other hand, the
relation of QMD to TDH shows that no quantum
correlations can be present beyond the mean-field level. These
questions demand more detailed discussions and are beyond the scope
of this section.

\subsection{Fluctuations}
\label{fluctuations}

Even though fluctuations are probably not important in the
simulations of the present code comparison, which looks only at
global one-body observables, they become critically important in
cluster formation. Fluctuations are also relevant to the question of
the initialization of a heavy-ion collision, which is treated in
more detail in Sec.~\ref{initialization}. Here we discuss in general
how fluctuations arise in the BUU and QMD approaches.

In BUU the TP simulation yields an exact solution of the BUU
equation in the limit of an infinite number of TPs. Thus BUU is a
deterministic equation, and in the above limit one obtains a unique
solution with no fluctuations. Fluctuations appear in practice,
because of the finite number of TPs and the stochastic simulation of
the collision term. These fluctuations are unphysical (but early on
they were used to gauge the most unstable mode of the
system~\cite{Colonna93}). However, from a physical point of view the
BUU transport theory should be extended to include fluctuations
since dissipation (collision term) should always be accompanied by
fluctuations~\cite{AbeAyik96}. Additional sources of fluctuation
arise from initial correlations and from the truncation of higher-order correlations. The inclusion of fluctuations leads to the
Boltzmann-Langevin equation with the addition of a fluctuation term
on the right-hand side of Eq.~(\ref{buueq})~\cite{AbeAyik96}.
Recently methods were devised to introduce fluctuations, from the
early BOB method (Brownian one-body fluctuation~\cite{BOB}), to the
SMF (stochastic mean field) formulation with density fluctuations~\cite{SMF}, and to the most recent BLOB (Boltzmann-Langevin one
body) approach~\cite{BLOB}. The latter approach implements fluctuations by a method modified from
the original proposal of Ref.~\cite{Bau87}, which moves with one TP collision a
swarm of $N_{TP}$ test particles, corresponding to a nucleon-nucleon
collision. In this sense it is closer to QMD with respect to the
collision statistics.

Due to its relation to molecular dynamics, QMD-type theories include
classical N-body correlations, and thus are expected to show more
fluctuations than BUU-type theories. The width parameter of the
Gaussians $\Delta x$, usually taken in the range
about $(\Delta x)^2 \sim 1-2$ fm$^2$, can be considered as a parameter to give a
reasonable amount of fluctuations and a good reproduction of the
surface properties of a nucleus. Note that QMD can be regarded as a
similar method to the parallel ensemble technique
   for BUU but without averaging over ensembles (or events) for the
   mean field and the Pauli blocking. As discussed above fluctuations will affect fragmentation in heavy-ion collisions. Indeed, a dedicated comparison of AMD and SMF actually showed a considerable difference in cluster yields~\cite{Rizzo07} with different fluctuation treatments in coordinate as well as in momentum space.

\section{Homework description}
\label{homework}

The goal of this paper is a comparison of results between codes but
not of code results to experiment. Here we compare the results
from different transport models under strictly controlled conditions. Controlled conditions imply not only the same physical
input, i.e., the same nuclear EoS and in-medium cross sections, but
also initial conditions of the reaction which are set up as close as
possible to each other, and are subject to the same requirements on the
accuracy of the simulation and the statistical significance. As in the
Trento 2009 workshop, Au+Au reactions at 100 and 400 AMeV have
been chosen as representative cases. From the
history of the code comparison and also for ease of reference,
calculations for the two different energies, 100 AMeV and 400 AMeV,
are referred as B- and D-mode respectively in this article.

In the homework, we specify the physics input precisely. The nuclear
EoS was specified as follows. In the case of non-relativistic
transport codes, the code practitioners were told to use the simple
Skyrme-type single-nucleon potential
\begin{equation}
U_{n/p} = \alpha\left(\frac{\rho}{\rho_0}\right) +
\beta\left(\frac{\rho}{\rho_0}\right)^\gamma \pm
2S_{pot}\left(\frac{\rho}{\rho_0}\right)\delta,
\end{equation}
where the parameters for the isoscalar potential are $\alpha=-209.2$
MeV, $\beta=156.4$ MeV, and $\gamma=1.35$, and the strength of the
symmetry potential at saturation density is $S_{pot}=18$ MeV. In the
above $\delta=(\rho_n-\rho_p)/\rho$ is the isospin asymmetry, and
sign $+$ is for neutrons and $-$ for protons. In the case of
relativistic transport codes, the code practitioners were told to
use the $NL\rho$ parameterization in the $\sigma\omega\rho$
relativistic mean-field model, with the values of the parameters from
Set I in Ref.~\cite{Liu02}. Both the non-relativistic and
relativistic nuclear EoS yields the same saturation properties of
nuclear matter, i.e., the saturation density of $\rho_0=0.16$ fm$^{-3}$,
the nuclear binding energy $E_0 = -16$ MeV, the incompressibility
$K_0 = 240$ MeV, and the symmetry energy $S(\rho_0) = 30.3$ MeV. A
Landau model with similar saturation characteristics is used for
pBUU~\cite{pBUU}.

For a better representation of nuclear surface in nuclei, many codes use a surface term, which is proportional to the density gradient or is represented by a Yukawa potential.
Since this introduces different behavior in the comparison of the
codes, which are difficult to ascertain, we asked the code practitioners
to turn off this option.

Regarding elementary processes, the code practitioners were told to
suppress inelastic processes and to use constant isotropic cross
section of 40 mb for elastic nucleon-nucleon scatterings.

The first step in a simulation is the specification of the
initialization of the colliding nuclei, which depends on the methods
chosen to represent the phase-space distribution, namely on point or
extended test particles in BUU or the shape of the nucleon wavepackets in QMD with or without antisymmetrization. As the simplest
prescription, the code practitioners were asked to follow an initial
density distribution of Woods-Saxon form
\begin{equation}
\rho(r) = \frac{\rho_0}{1+\exp[(r-R)/a]},
\end{equation}
where $\rho_0 = 0.16$ fm$^{-3}$ is the saturation density, $R =
1.12A^{1/3}$ (fm) is the nuclear radius, with $A = 197$ for Au
nucleus, and $a = 0.6$ fm is the diffuseness parameter. As discussed
later, it is not easy to obtain an exactly identical initialization,
thus we asked the code practitioners to map their initial density
distributions as closely as possible using the type of test particle
or particle distribution in their codes. Once the coordinate space
is sampled, the local density approximation is used to sample the
momentum of each nucleon, i.e., the momentum of each (test) particle
should be chosen randomly in the local Fermi sphere according to the
local density, generally isospin-dependently. The initial distance
between the centers of the projectile nucleus and the target nucleus
should be set to 16 fm in the beam direction.

The representation of the phase space is different in BUU- and
QMD-type models, which also affects the statistical significance of
the calculations. To make these approximately comparable, we made
the following specifications: 100 test particles per nucleon and 10
simulations with different initializations for BUU models; 1000
simulations with different initializations for QMD models. Since it
is difficult to generate 1000 stable initial configurations for QMD
models, sometimes they have been generated by rotating a few stable
initializations randomly in space. The time evolution of the
simulations is followed until 140 fm/c at 100 AMeV and 100 fm/c at
400 AMeV. The time step in the evolution is left to the code
practitioners, but is recommended to be set as 0.5 fm/c.

Transport calculations basically have two main ingredients, the mean
field (related to EoS), and the NN collisions, which have different
influence on the reaction dynamics. It is important to understand
the influence of these ingredients in the codes. Therefore, in
addition to the full calculations that include both the mean field
and NN collisions, we also asked for B-Vlasov calculations, using
only the mean field and turning off collisions, and for B-Cascade
calculations, with only collisions and no mean field. Note that B
mode refers to simulations at the incident energy of 100 AMeV.

For the output of the calculations we asked for the full information
from the code practitioners, to make it possible to generate the
observables in the same way for all codes and also to be able to
inspect additional quantities later on. A transport calculation has
the advantage that one may look into the collision at any stage of
the evolution, something one cannot do in experiments. The code
practitioners provided two types of files: (test) particle files
which specify the type, position, and momentum of each (test)
particle at times 0, 20, 40, 60, 80, 100, 120, and 140 fm/c for each
simulation; collision files, which specifies the type, (effective)
mass, time, and momentum of the two colliding (test) particles for
each attempted collision, and the result of the Pauli blocking
(successful or not). From these files we generated several
quantities which are discussed in the next two sections.

\section{Initial Configurations and Stability}
\label{initialization}

Ideally all codes should start with the same initial configurations
so that one can disentangle the effects of the initial conditions
and the reaction dynamics. Although the initial density
distributions of a Woods-Saxon form was detailed in the homework, as
described in Sec.~\ref{homework}, this procedure for the
initialization turned out to be not quite satisfactory, since it
does not guarantee that the initial nuclei are really in the ground
state corresponding to the nuclear mean field chosen in the
homework. To check the stability of the initial configurations,
simulations were first performed at a large impact parameter,
$\text{b}=20$ fm, so that the projectile and target were far enough
apart that essentially no nucleons or energy should be exchanged.
The mean density evolution of an Au nucleus obtained from an average
of the projectile and target nuclei was then used to check the
stability of the initial configurations.

\subsection{Initial Configurations}
\label{initializationa}

\begin{figure}[ht]
\includegraphics[scale=1]{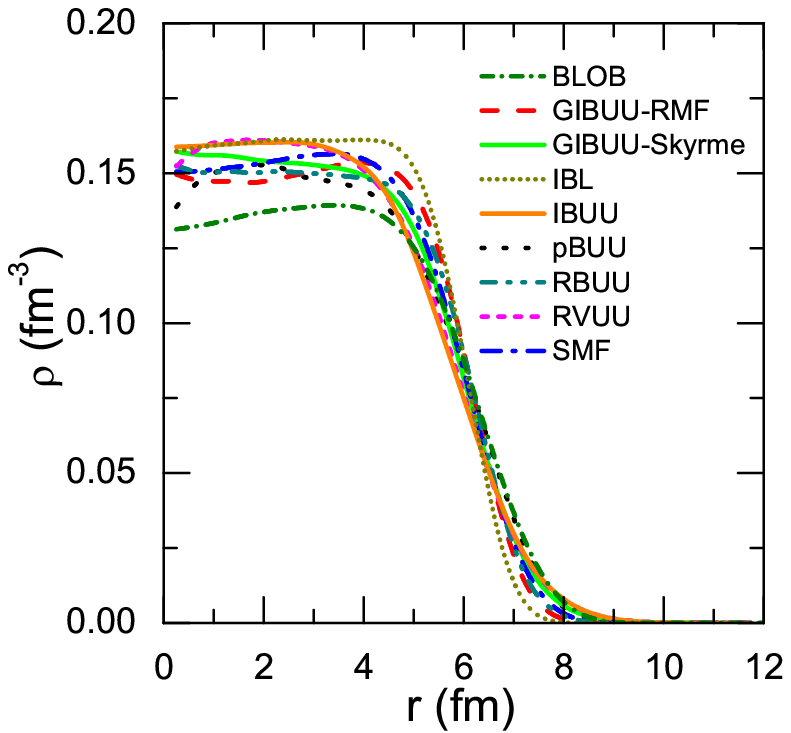}
\includegraphics[scale=1]{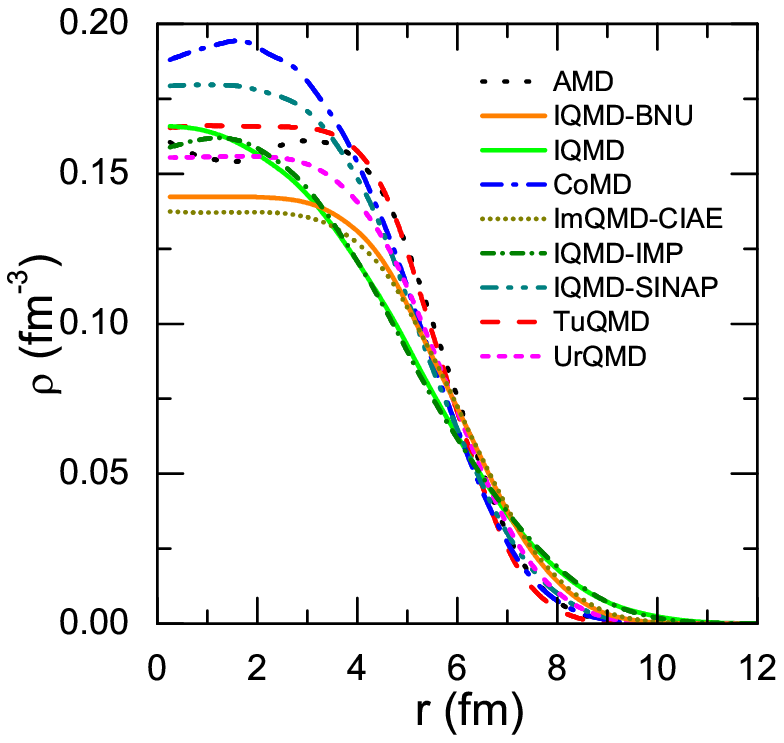}
\caption{(Color online) Initial density profiles for BUU-type (left
panel) and QMD-type (right panel) models. 
} \label{denini}
\end{figure}

There are many treatments for the initialization of the nuclei. For
instance, nucleons may be treated as point particles, Gaussian-type
finite-size particles~\cite{Har89,Aic91}, or triangular-type
finite-size particles~\cite{BLOB,SMF}, and in QMD models the width
of Gaussian wavepacket depends on the size of the collision
system~\cite{Aic91,Wan02}. Generally, BUU-type models use a given
density distribution provided by a Woods-Saxon parameterization, the
Skyrme-Hartree-Fock calculation~\cite{Vau72}, or the Thomas-Fermi
approximation~\cite{pBUU,TF2}. For QMD-type models, the initial
configurations of nuclei are usually selected in such a way that
they yield similar binding energies and charge radii as the
experimental data. Sometimes a minimum distance between two
arbitrary nucleons is required to give a more uniform initial
phase-space distribution (IBL, IQMD-BNU, ImQMD). In some models
(pBUU, CoMD, IQMD-SINAP), a frictional cooling method is used to
reach a ground-state initialization. More information about the
initialization of different codes used in the present work is given in
Table.~\ref{apt1} and in the references listed in Table.~\ref{T1}.

The density profiles at $t = 0$ fm/c from BUU-type models and
QMD-type models are shown in Fig.~\ref{denini}. These are averaged
density distributions of the projectile and target over all the
calculations requested in the homework (i.e., 10 runs with 100 TPs
for BUU and 1000 events for QMD). The density distributions are
obtained by folding the test particle positions with the width of
the (test) particle as provided by the code practitioners, or by a default
value of 1.5 fm. 
The initial phase-space distributions are seen to vary from
code to code. BUU-type models (left panel) have a smaller
dispersion, while QMD-type models (right panel) are
more different from each other. AMD in its more quantum nature
succeeds very well to produce good ground states and resembles the
closest the Woods-Saxon distribution profile. AMD also exhibits an
oscillatory behavior in the center similar to shell effects observed
in Hartree-Fock calculations. The representation in
Fig.~\ref{denini} emphasizes the differences in the interior, while
there are also considerable differences in the tail. To show this,
the deviation values of $<r^2>^{1/2}$ and $<r^4>^{1/4}$ from that of
the required Woods-Saxon ditribution for each code are listed in
Column 4 and 5 in Table~\ref{apt1}. The main difficulty in
reproducing the given Woods-Saxon density profile is that the
initialization depends on the implementation of individual codes and
the shape of the test particles or wavepackets. Of course,
differences in the initial momentum distributions are also
important, especially when comparing observables related to the
final momentum distributions, as in Sec.~\ref{secy} and \ref{secv1}.
Initial momentum distributions are shown later (upper panels of
Fig.~\ref{rap_B}).

From the above comparison, we learnt the different situations in the initializations between BUU and QMD approaches. In BUU, with the quasi-continuous distribution function for many
test particles, it is not too difficult to obtain initial states
with a reasonably smooth density and momentum profile. These initial
states are usually also more stable, especially if they are
calculated in a Thomas-Fermi or Hartree calculation using the same
energy-density functionals as in the transport calculation, rather
than prescribed as a density profile. For QMD approaches, with their
larger fluctuations, the initially prepared states are less stable
and many attempts are necessary to find stable initial states. Often
criteria on the binding energy of the initialized nuclei and/or
stochastic cooling are used additionally.

\subsection{Density Evolution}

\begin{figure}[h]
\includegraphics[scale=1.65]{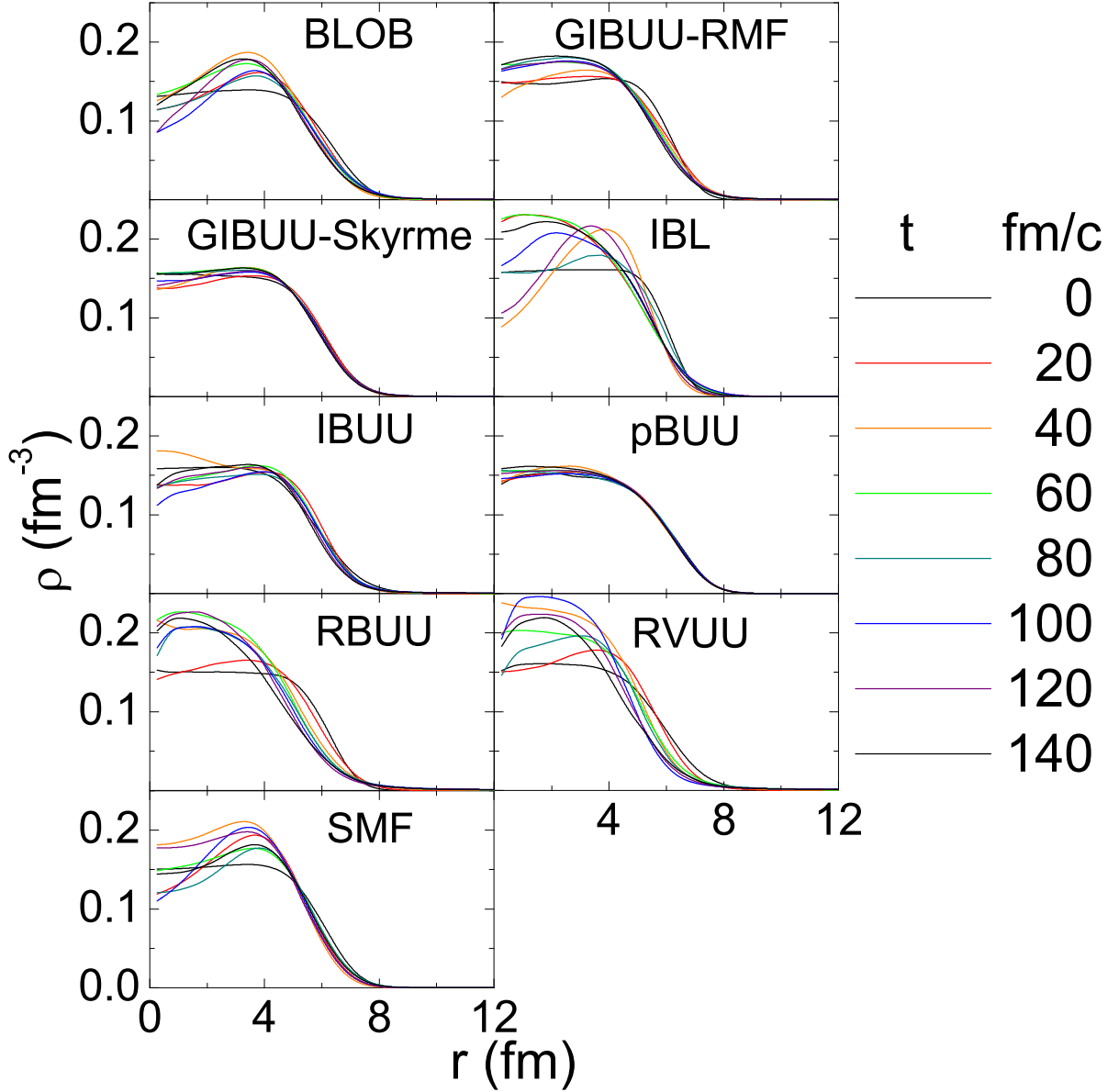}
\caption{(Color online) Time evolution of the density profiles
in a single Au nucleus in steps of 20 fm/c (see legend for the
explanation of different color lines) for BUU-type models at
$\text{b}=20$ fm. }\label{stability_buu}
\end{figure}

\begin{figure}[h]
\includegraphics[scale=1.65]{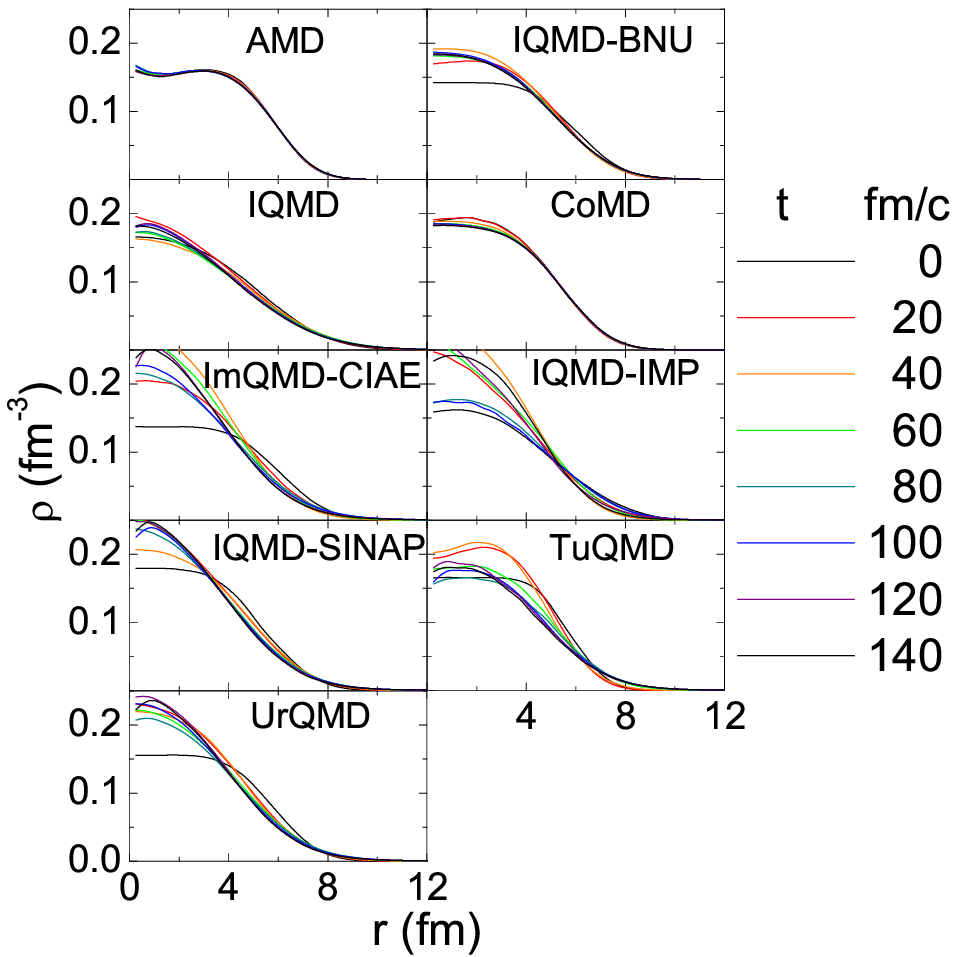}
\caption{(Color online) Same as Fig.~\ref{stability_buu} but for QMD-type models.}
\label{stability_qmd}
\end{figure}

To check the stability of the initial configurations as described in the previous subsection, we examine the time evolution of radial density distributions of nuclei every 20 fm/c.
Figures \ref{stability_buu} and \ref{stability_qmd} show the
corresponding density profile as a function of radial distance for
BUU-type and QMD-type models, respectively. Ideally, if the initial
nuclei are stable, there should be no time dependence of the density
distributions. Examples are GIBUU-Skyrme, pBUU, AMD, and CoMD codes,
where the density profiles show stability within the simulation time
of 140 fm/c. Generally, this requires special treatments such as
phase-space constraint or extremely good Pauli blocking, because the
transport models are not designed to calculate good ground states of
a nucleus. A bubble-like configuration is formed initially in the
IBL and possibly in the BLOB codes with the choice of the parameter
set imposed by the homework. Such structure is not evident in any
QMD codes. In other codes such as IBUU, RVUU, SMF, TuQMD, and
IQMD-IMP, the radius of the nucleus oscillates like a giant
monopole resonance. This is understandable as the given initial
density distribution may not represent the ground state. In many QMD
codes (IQMD-BNU, IQMD-IMP, IQMD-SINAP, and UrQMD), the nucleus evolves
away from the initial density distribution (black lines) quickly and
most relax into a reasonably stable configuration. Such instability
seen for the initialized nuclei demonstrates the difficulty of
imposing a common initial configuration to different codes. The different initial density distributions from all 18 codes shown
in Fig.~\ref{denini} illustrate that a common initialization in a
code comparison is more involved than naively expected. The
representation of the system with nucleons or with a
quasi-continuous distribution function and the shape of the (test)
particles are closely connected and affect the evolution and
stability. The procedure chosen here to prescribe a common density
profile of the initial nuclei is not optimal and often produces an
initial configuration that is not stable. It may be more important to
start with reasonably good ground states such that no spurious
evolution affects the results. In this respect box calculations as
discussed in Sec.~\ref{summary} for the future of this code
comparison project may avoid this difficulty of the initialization.

\section{Heavy-ion collisions at $\text{b}=7$ fm} \label{results}

In this section we present and discuss the comparison of the
different transport codes when employed to simulate a heavy-ion
collision. The impact parameter was chosen to be 7 fm, which
corresponds to a reduced impact parameter of about 0.5. In such a
reaction violent interactions take place and all aspects of a
heavy-ion simulation are going to be important: the initialization
of the system, the mean-field propagation, and the collision term
with the collision probabilities and the Pauli blocking, thus
allowing us to observe and understand the model dependencies. In
doing so, one should also be able to observe many features of
heavy-ion collisions, especially since in a simulation we have the
advantage of being able to look into every aspect of the evolution.

We will look into the different aspects of a collision. In the first
subsection, we discuss the density evolution of the reaction
qualitatively in contour plots. In the second subsection, we look in
detail into the action of the collision term with respect to two
aspects, the probability of a NN scattering and the effect of Pauli
blocking. We then discuss observables, which are commonly used in
the comparison of a simulation with experiment, the rapidity
distribution and the collective flow. The results are the average from all runs of
the homework, and we do not show the statistical width of the results of
each code, in order not to make the figures too crowded. As before, we group together
BUU and QMD codes, respectively, in the figures. One expects and
does in fact see that there are many similarities between the
results of these two groups, but one also sees characteristic
differences, associated with the different strategies of these codes
as discussed in Sec.~\ref{theory}.


\subsection{Density Evolution}

\begin{figure}[h]
\includegraphics[scale=1.4]{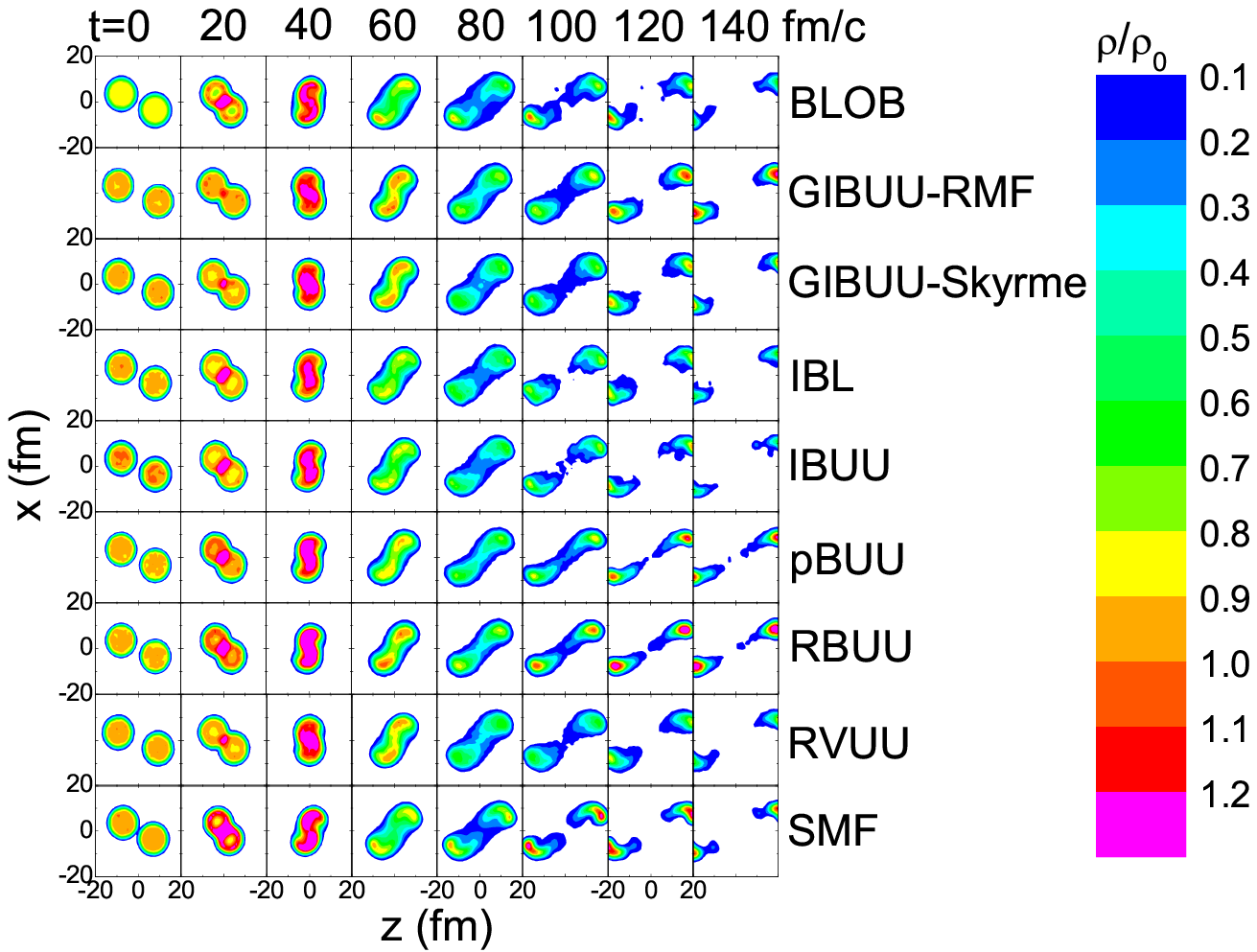}
\caption{(Color online) Average density contours in steps of 20 fm/c in Au+Au
collisions at impact parameter $\text{b}=7$ fm and beam energy 100
AMeV from BUU-type models. This is for the B-Full mode that includes both
mean-field potentials and nucleon-nucleon scatterings.}
\label{denevo_buu}
\end{figure}

\begin{figure}[h]
\includegraphics[scale=1.4]{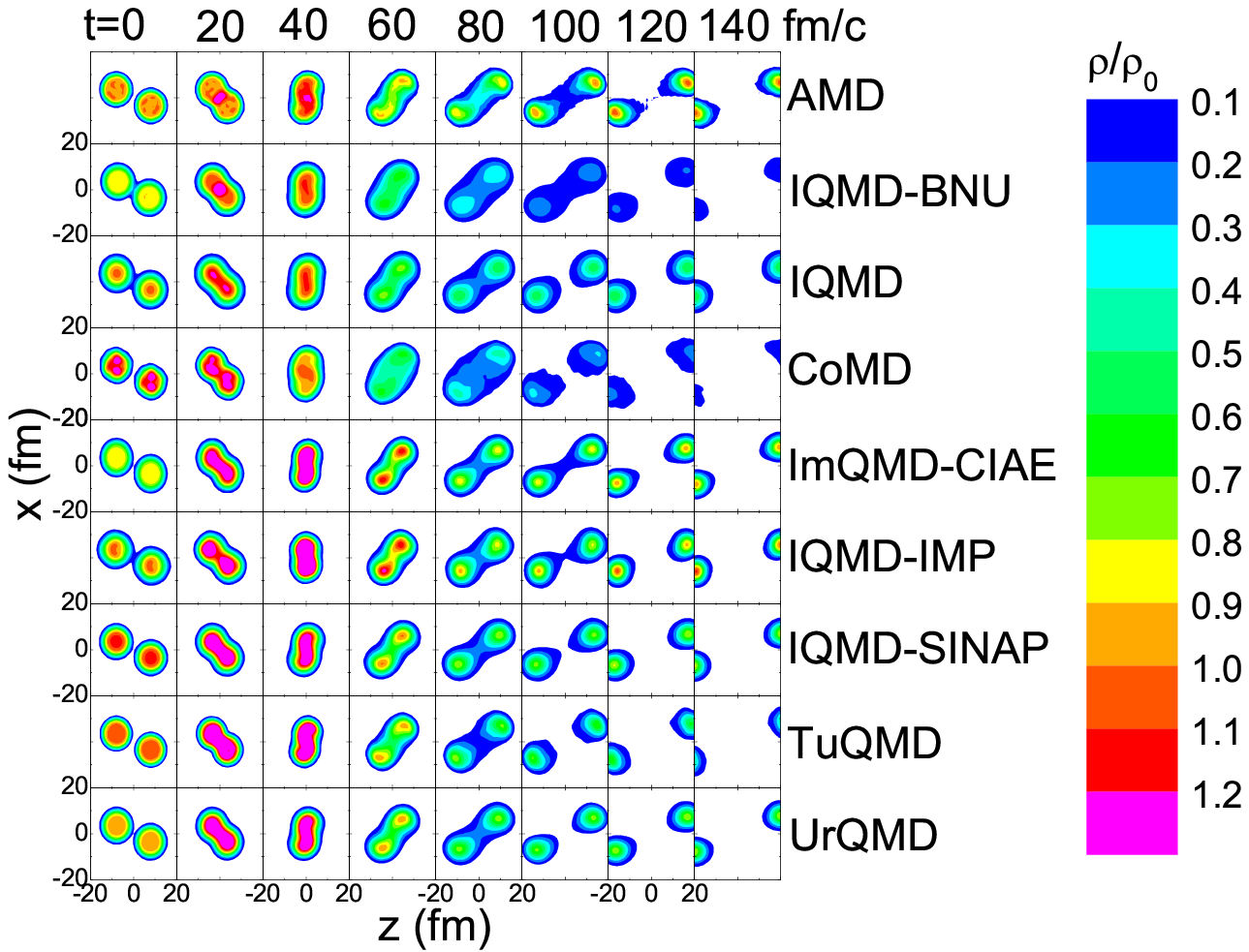}
\caption{(Color online) Same as Fig.~\ref{denevo_buu} but from
QMD-type models.} \label{denevo_qmd}
\end{figure}

\begin{figure}[ht]
\includegraphics[scale=0.8]{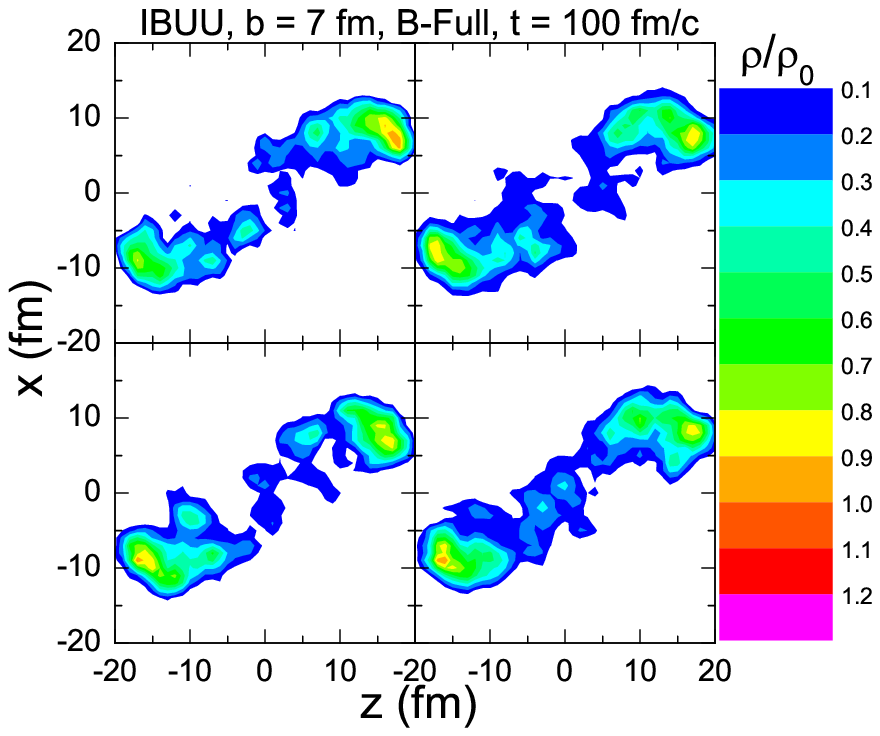}
\includegraphics[scale=0.8]{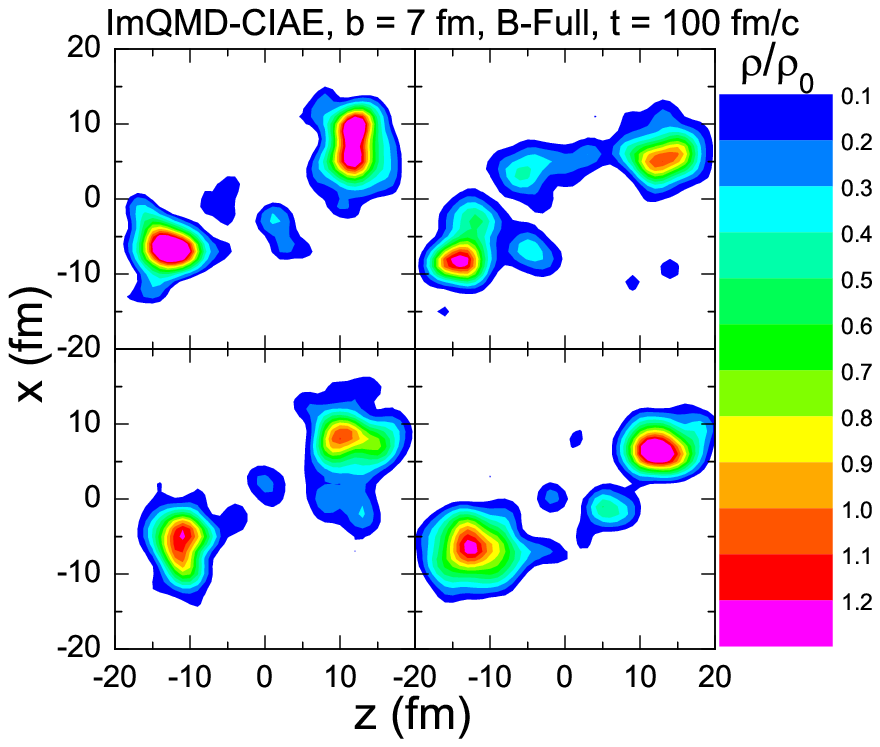}
\caption{(Color online) Density contours for 4 runs with 100 TPs per nucleon from IBUU (left) and four individual events from ImQMD-CIAE (right) out of the collisions displayed in Figs. \ref{denevo_buu} and \ref{denevo_qmd} at $t=100$ fm/c. } \label{den_indiv}
\end{figure}

The density contours in the $x$-$o$-$z$ plane (i.e., the reaction plane
with $x$ the direction of the impact parameter and $z$ the beam
direction) in steps of 20 fm/c in Au+Au collisions at 100 AMeV are
displayed in Figs.~\ref{denevo_buu} and \ref{denevo_qmd} for the BUU
and QMD models, respectively. The contour plots give a good
qualitative impression of the dynamical evolution in different
models. The plots actually represent the average over all runs,
i.e., 10 runs with 100 TPs in BUU and 1000 runs ("events") in QMD.
This averaging smears out fluctuations, which are expected to be
stronger in QMD models than in BUU models.

The general progression of a heavy-ion collision is exhibited in all
models: the merging and maximum compression up to about 40 fm/c, the
development of sideward flow from about 60 to 80 fm/c, and the
formation and subsequent breaking of a neck at about 100 fm/c. From
100 fm/c onward, one observes the formation and evolution of the
projectile- and target-like residues, which are clearly highly
excited and develop their own dynamics. The final de-excitation of
the excited fragments is not the subject of this comparison, but should be taken into account, if one wants to
compare to experiment. In this collision, one could consider 140
fm/c as the freeze-out time, after which the de-excitation of the
primary fragments is usually calculated with a statistical code.

Consistent with the density profiles of Fig.~\ref{denini}, we
clearly see in the contour plots of Figs.~\ref{denevo_buu} and
\ref{denevo_qmd}, that the initial states in the different codes are
rather different with more or less steep density profiles. It is
interesting to see that higher initial densities, which occur more
often in QMD codes, lead to an earlier and stronger development of
the high-density phase in a reaction. This supports the
suggestion that differences in the initialization may actually lead
to differences in the evolution and the physics observables and
underscores the difficulty of code comparisons without identical
initial states. The maximum density is reached by
all models around 40 fm/c. The time span from 20 to 40 fm/c is very
important for many observables, such as stopping, flow, and particle
production, since in this interval most of the elementary collisions
occur, as will be discussed in the next subsection.

One can see that here the evolution is characteristically different
in BUU and QMD models. The density pattern is more detailed in BUU
with distinctly compressed central zone and often normal density at
the center of the residues. In QMD the density pattern appears more
uniform, but this is primarily due to the averaging over many
events. To demonstrate this, we show in Fig.~\ref{den_indiv} the
density contour plots at the time of 100 fm/c for four different
runs each generated by one BUU code (IBUU) with 100 TPs and by one
QMD code (ImQMD-CIAE). The final states in the QMD model show a large amount
of fragmentation. The difference in fragmentation patterns in
Fig.~\ref{den_indiv} between BUU (left panels) and QMD (right
panels) also reflects to which extent small structures can be
generated by the different representation of the phase space by
test particles or nucleons. In BUU these fine structures partly
survive in the averaged distributions of Fig.~\ref{denevo_buu}. The
similarity of the four runs also demonstrates the more deterministic
nature of the BUU method. As discussed in Sec.~\ref{fluctuations}, without introducing additional stochastic mechanisms (such as in BLOB or SMF), the fluctuation in BUU-type models depends on the test-particle numbers, and with infinite number of test particles the simulation leads to a deterministic solution of the BUU equation. For QMD-type models, the fragmentation character can be related to the width of the Gaussian wavepacket, and they are expected to show more fluctuations than in BUU-type models. A detailed study of the
fluctuation and fragmentation patterns of different codes is planned
in future efforts of this code comparison, as will be discussed below in the
Outlook. It is thought that for the bulk one-body observables discussed
here these will be of lesser importance, but they may already have an
effect on the collision rates.

There are also differences in the evolution of the neck between 60
and 100 fm/c. In BUU the neck is usually fatter and stretches out
farther. In the breaking of the neck fine structures (or even small
fragments) appear. The residues are strongly deformed for a long
time. In QMD the neck breaks faster and the residues rather quickly
approach a spherical shape. Again these differences are mostly due
to the averaging, since in single QMD events fragments are formed in
the neck as shown in the right panels of Fig.~\ref{den_indiv}.


Considering these differences, it will not be surprising that
different codes will show differences in the collisional
characteristics to be discussed in the next subsection, and in observables to be
discussed in subsection C.

\subsection{Collision term}

\begin{figure}[h]
\includegraphics[scale=1.25,bb = 0 0 421 297]{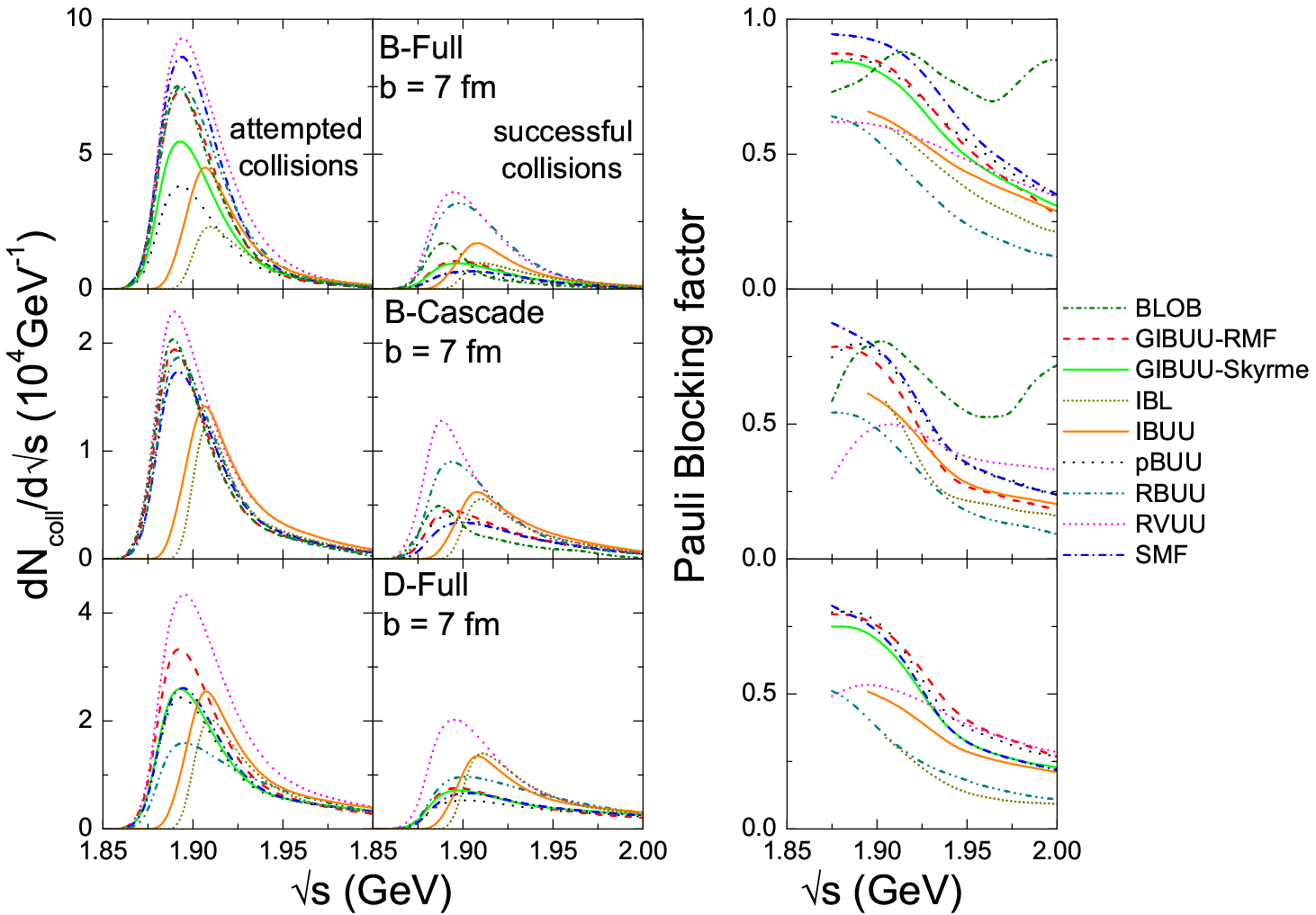}
\caption{(Color online) Center-of-mass energy dependence of
nucleon-nucleon attempted (left) and successful (middle) scattering
numbers as well as the Pauli blocking factors (right) for the B-Full
mode (top), the B-Cascade mode (middle) both at 100 AMeV, and the
D-Full (bottom) mode at 400 AMeV at impact parameter $\text{b}=7$ fm
from BUU-type models.} \label{colls_buu_b7}
\end{figure}

\begin{figure}[h]
\includegraphics[scale=1.25]{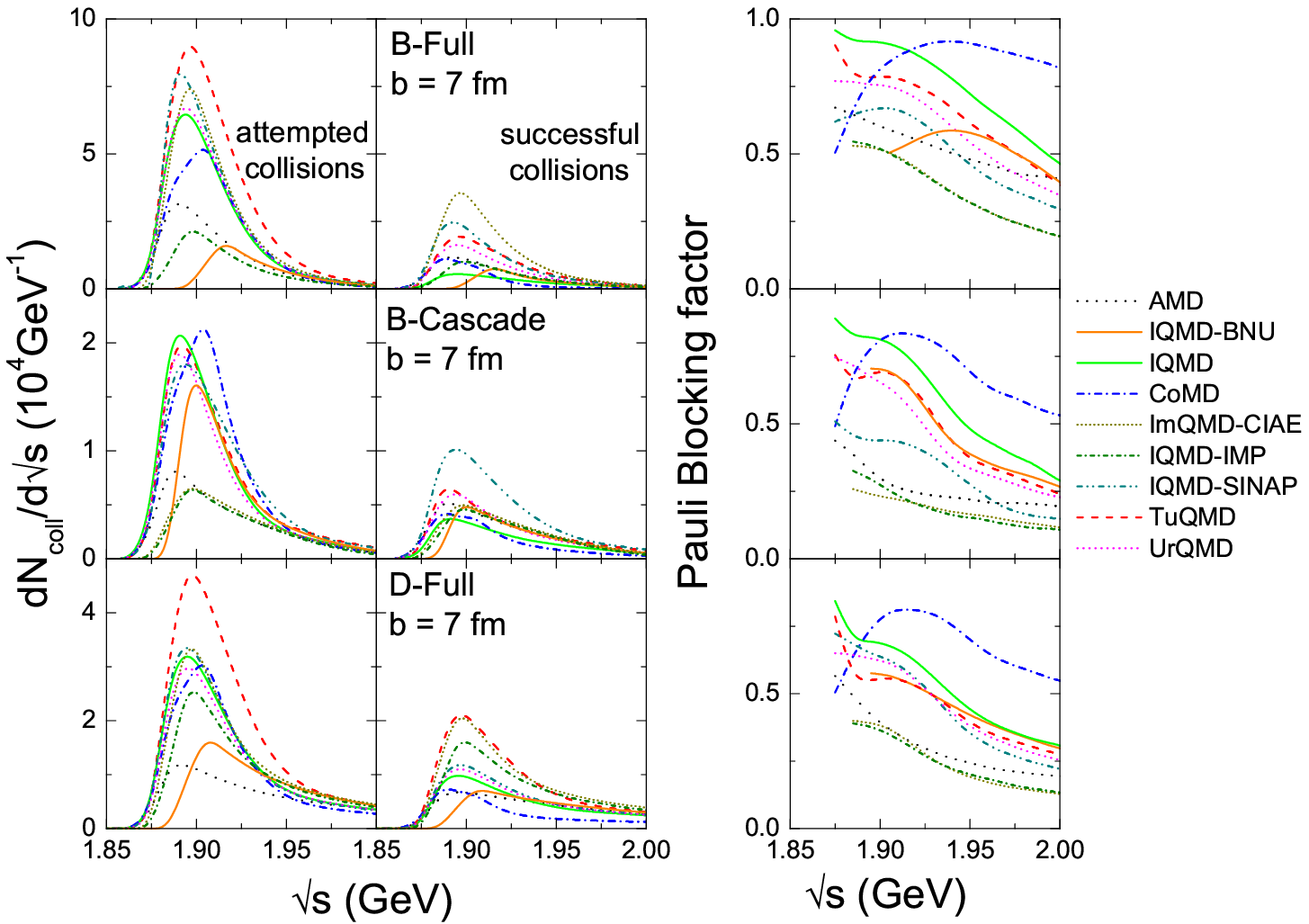}
\caption{(Color online) Same as Fig.~\ref{colls_buu_b7} but from
QMD-type models.} \label{colls_qmd_b7}
\end{figure}

The collision term is the other important ingredient besides the mean
field. As formulated in Eq.~(\ref{NNsca}), it is highly non-linear
in distributions and non-local in momentum, and therefore is
simulated stochastically, as discussed in Sec.~\ref{theory}. The
collision term is crucial for the evolution of the simulation of a
heavy-ion reaction, since it is the cause for energy dissipation. It
is also the part of a simulation where different codes differ most
in the implementation. It is therefore worthwhile to examine the
collision rates in detail, even though these are not observables.

We have studied both the time evolution of the total collision rates
and their distributions over energy. We show only the latter here,
since they provide better physics insights. We note that the time
evolution of the collision rates is closely linked to the density
evolution, with the highest rates in the densest phases. They
therefore also trace the density evolution shown in
Figs.~\ref{denevo_buu} and \ref{denevo_qmd}. For instance, one
observes rise and fall in the collision rates when the residues
oscillate.

The number of collisions per 100 keV bin are plotted for BUU and QMD
codes in Figs.~\ref{colls_buu_b7} and \ref{colls_qmd_b7},
respectively, as a function of the center-of-mass (CM) energy
$\sqrt{s}$ of the individual nucleon-nucleon collision. The
arrangement of both figures is the same. In the left column we show
the "attempted" collisions, which according to the definition in
Sec.~\ref{theory} are the collisions where the distance criterion
(and in some cases additional criteria) are satisfied. The middle
column shows the "successful" collisions, which are the attempted
collisions where the final state is not Pauli-blocked. The rightmost
column shows the Pauli blocking factor, defined as
$1-(successful/attempted)$. The upper and bottom rows show the
B-Full and D-Full mode, respectively, i.e., a simulation with mean
field and collisions at 100 and 400 AMeV, respectively. The middle
row shows the B-Cascade mode, i.e., a simulation without the action
of the mean field. Obviously we do not show the B-Vlasov mode, since
there are no collisions there.

Generally we see the following behavior: in the NN frame we have the
relation $s=4(m^{*2}+p^2)$, where $m^*$ is the effective mass and
$p$ is the modulus of the 3-momentum in the CM frame of the NN
collision. The threshold with the prescribed free nucleon mass of
938 MeV is $\sqrt{s}=1.876$ GeV. In the figure this sharp cutoff is
slightly smeared out due to the plotting procedure. For relativistic
codes the effective Dirac mass and dressed momentum provided by the
code authors were used in the analysis, compared with the results
from the bare mass and momentum for non-relativistic codes.

The $\sqrt{s}$ for a free NN collision is 1.925 and 2.066 GeV for an
incident energy of 100 and 400 AMeV, respectively, and 1.894 GeV for
particles at the Fermi momentum. In Figs.~\ref{colls_buu_b7} and
\ref{colls_qmd_b7} it is seen that the peak energy for the maximum collision number is only
approximately 1.9 GeV for both incident energies. By checking also the time evolution of the NN scattering number, we found that most of the collisions occur in the most compressed stage, when most of the nucleons are stopped
to a large degree. However, the energy distribution has
a long tail at higher CM energies, which is even more extended for 400
AMeV, as expected.

The number of collisions is considerably smaller in the Cascade
mode, because here the nuclei disintegrate faster, due to the lack
of mean field, resulting in lower densities. At 400 AMeV, the number is even
smaller because of the faster disintegration of the
system. Note the change of scale for the number of attempted
collisions for the B-Cascade and D-Full modes.

The number of successful collisions is obviously lower than that of the
attempted collisions. The shape of the energy distribution of the
successful collisions looks similar to that of the attempted
collisions, but this, in fact, is not quite so as shown by the
blocking factors in the third column, which are not constant. The
blocking factors are larger at lower energies, since the final
states of softer collisions are more likely to be blocked due to the
smaller phase space. The blocking is reduced for higher NN collision
energies, since more of the phase space is free, particularly for
the higher incident energy.

There are considerable differences in the implementation of the
collision term in the different codes (see Table~\ref{apt2}).
In some codes (IBL, IBUU, and IQMD-BNU) collisions start with a higher
threshold than the free one. Here threshold is introduced to
suppress very soft collisions, based on the argument that these are
often spurious. In these codes the energy distributions are
different from those without artificial collision threshold.

The distributions of the attempted collisions in the B-Cascade mode
are rather similar between many codes both for BUU and QMD (but not
so much for the successful collisions). This is what one expects as
a consequence of no mean-field dynamics and the use of the same
cross sections. Exceptions are the codes IBL, IBUU, AMD, ImQMD-CIAE, and
IQMD-IMP probably due to different treatments of collision
threshold or due to different initial density distributions.

The successful collisions are more important quantities in the
Full modes, because they determine the energy dissipation. The
collision numbers are higher in some cases for the relativistic
codes (RBUU at 100 AMeV, RVUU at both energies, TuQMD at 400 AMeV,
but not GIBUU-RMF) and also for IQMD-SINAP and ImQMD-CIAE. The peak
of the energy distribution is shifted in some cases: to the lower
side for BLOB, to the higher side for IBL, IBUU, and IQMD-BNU. In
pBUU and IQMD the total number of successful collisions is
relatively low, and also for AMD and CoMD at the higher incident
energy. The last observation is understood for AMD which treats a NN
collision as that of two phase-space wavepackets. Since the
collision probability is proportional to the relative velocity
between the centroids without velocity fluctuation, the number of
collisions is smaller than those in other models at lower
$\sqrt{s}$. A similar situation might exist for CoMD, where
particularly the collision numbers are much smaller at higher
$\sqrt{s}$. A contrary case exists for the relativistic QMD code
TuQMD, where the number of collisions is typical at 100 AMeV, but
higher at 400 AMeV.

The Pauli blocking factors in the third column show that, to what
extent the differences in the successful collisions are due to the
blocking. Generally the blocking factors are considerably different
between codes but the trends are similar, i.e., they have similar
shapes for different codes, signifying that the blocking behavior
does not depend very much on the CM energy of the NN scattering. The
blocking factors tend to converge better at the higher NN scattering
energies, especially at the higher incident energy of 400 AMeV (with
BLOB data not available). This is expected since in high-energy
collisions the density evolution of the collision matters less.
There are some exceptions in the blocking factors. CoMD has a very
different energy dependence of the blocking factor. This behavior
reflects phase-space correlations produced in the initial
configurations, which inhibit collisions with relative momenta near
the Fermi momentum (inner particles), but not so much at the surface
(low relative momenta) because of lower densities.

In summary, there are considerable differences in the behavior of
the collision terms between different codes, which, however, tend to
diminish at higher incident energies. Aside from the different
initializations as discussed earlier, these differences or
similarities are most likely behind the differences and similarities
for the behavior of the observables, which will be discussed in the next
two subsections.

\subsection{Rapidity distributions}
\label{secy}

\begin{figure}[t]
\includegraphics[scale=1.5]{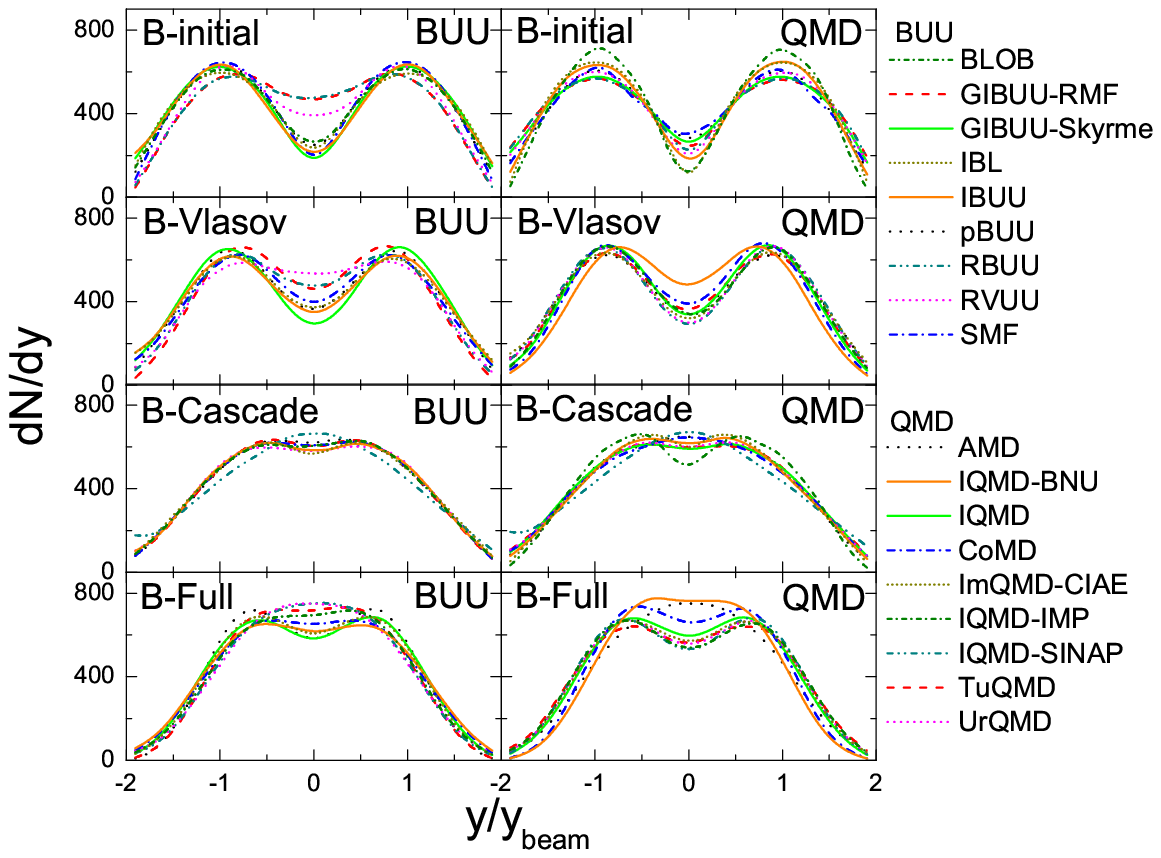}
\caption{(Color online) Initial rapidity distributions (top panels) at the beam energy of 100 AMeV and final
rapidity distributions for the B-Vlasov mode (next panels down), the
B-Cascade mode (next panels down), and the B-Full model (bottom
panels) at impact parameter $\text{b}=7$ fm from BUU-type (left
panels) and QMD-type (right panels) models. } \label{rap_B}
\end{figure}

\begin{figure}[t]
\includegraphics[scale=1.4]{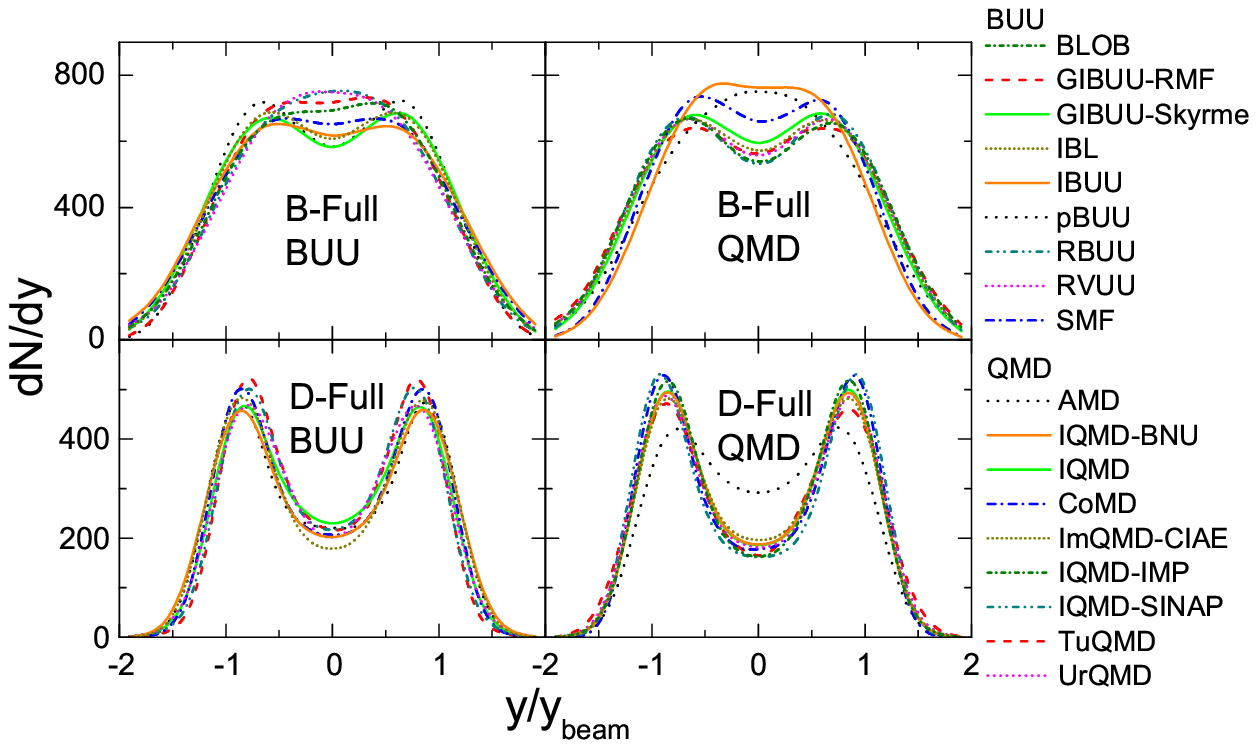}
\caption{(Color online) Rapidity distributions at $\text{b}=7$ fm for the B-Full (top panels) and D-Full (bottom panels) mode from BUU-type (left panels)
and QMD-type (right panels) models.} \label{rap_BD}
\end{figure}

A rapidity distribution characterizes the distribution of particles
along the beam axis in a Lorentz invariant manner. The distributions
are usually displayed relative to the beam rapidity, so that
projectile and target sit at $y/y_{beam}=\pm1$ in the CM
frame, respectively. A rapidity distribution tests the stopping of
the nucleons in a heavy-ion collision and is an important basic
observable.

In Fig.~\ref{rap_B} we show the rapidity distributions for the
B-mode (100 AMeV) calculation from BUU codes on the left and from QMD codes on the
right. The first row shows the initial distributions and the next
rows are the final distributions for the Vlasov, Cascade, and Full modes,
respectively. In
Fig.~\ref{rap_BD}  we compare the results of the Full mode for 100
and 400 AMeV incident energy. The initial distributions exhibit a
double-humped structure. These are the expected superposition of the initial
target and projectile distributions boosted to $\pm$beam rapidity as mentioned.
In the Vlasov mode, the peaks in final distributions are moved
somewhat inward to midrapidity, corresponding to a braking of the
longitudinal velocity due to the action of the Coulomb and nuclear
mean fields, and to the collective deflection of the motion in
transverse direction (see the next subsection where this is shown in
more details). The NN scatterings lead to a filling of the
midrapidity distribution in the B-Cascade mode, since they convert
the longitudinal into random momentum (especially for an isotropic
cross section used here). The amount of filling, called stopping and
the opposite behavior called transparency, should depend on the NN
scatterings. The Full mode combines both effects. Generally the
peaks of the Full mode are between those of the Vlasov and Cascade
distributions, but the exact shape can be different depending on the
code. In Fig.~\ref{rap_BD} one sees that the stopping is weaker at
the higher incident energy, correlating with the smaller number of
collisions, already seen in Figs.~\ref{colls_buu_b7} and
\ref{colls_qmd_b7}.

As shown in Fig.~\ref{denini}, the initial rapidity distributions
for all BUU codes are rather similar, as was intended. We note that
for the relativistic BUU codes (GIBUU-RMF, RBUU, and RVUU), the Dirac
effective mass and the dressed momentum provided by the code practitioners
are used to calculate the rapidity, which leads to a flatter
rapidity distribution compared with those from non-relativistic
codes.  There are, however, larger differences in the QMD codes,
correlating with the larger spread in the initial density profiles
seen in Fig.~\ref{denini}.

One would expect that the rapidity distribution for the B-Vlasov
mode should be similar for all the codes, but this is not completely
true as shown in the second row of Fig.~\ref{rap_B}, most likely due to the
different initial density and momentum distributions.

In the B-Cascade mode the results are remarkably similar for all BUU
codes, but larger differences appear for the QMD codes. One can
relate these differences to those in the successful collisions in
the Cascade mode in Figs.~\ref{colls_buu_b7} and \ref{colls_qmd_b7},
which explains some (but not all) of the different behaviors. The
large stopping in the RBUU code is correlated with the large
collision numbers in Fig.~\ref{colls_buu_b7} (but this is not the
case for the large collision numbers for RVUU). For the IQMD-SINAP
code there is a clear correlation between the almost complete
stopping with the large number of collisions. For AMD and CoMD the
stopping is relatively strong, even though the collision numbers are
not particularly high. For IQMD-IMP the larger transparency is
related to the somewhat lower collision numbers.

In the Full mode we see the competition between the mean field and the NN scatterings. In
BUU models the rather similar Cascade distribution are split up by
the more different Vlasov results, and the other way around for QMD
codes. For the Full mode in both types of codes there are
considerable differences. There is again a correspondence between
the stopping and the number of successful collisions in the Full
mode. The relativistic BUU codes have stronger stopping, which might
be due to their initial rapidity distribution. For CoMD the strong
stopping from the full calculation is related to its initial
rapidity distribution, while for IQMD-BNU the large deflection in the
Vlasov mode seems to dominate also in the Full mode.

In Fig.~\ref{rap_BD} the similarity of the rapidity distribution at
the higher incident energy of 400 AMeV for almost all codes is
remarkable. In the case of AMD, probably too many nucleons
participate in violent collisions, due to the insufficient precision
of the physical-coordinate representation discussed in Ref.~\cite{AMD},
which overestimates the radius of the Au nucleus. The similarity of
the rapidity distributions at 400 AMeV appears in spite of the
differences in the collision spectra. The reason may be that the
stopping at higher incident energies is dominated by the NN
scatterings and particularly those at higher CM energies, where
almost all codes converge rather well, as seen in
Figs.~\ref{colls_buu_b7} and \ref{colls_qmd_b7}. Thus stopping is a
rather robust observable between different codes at high
incident energies.

\subsection{Anisotropic collective flow}
\label{secv1}

\begin{figure}[h]
\includegraphics[scale=1.5,bb = 0 0 340 234]{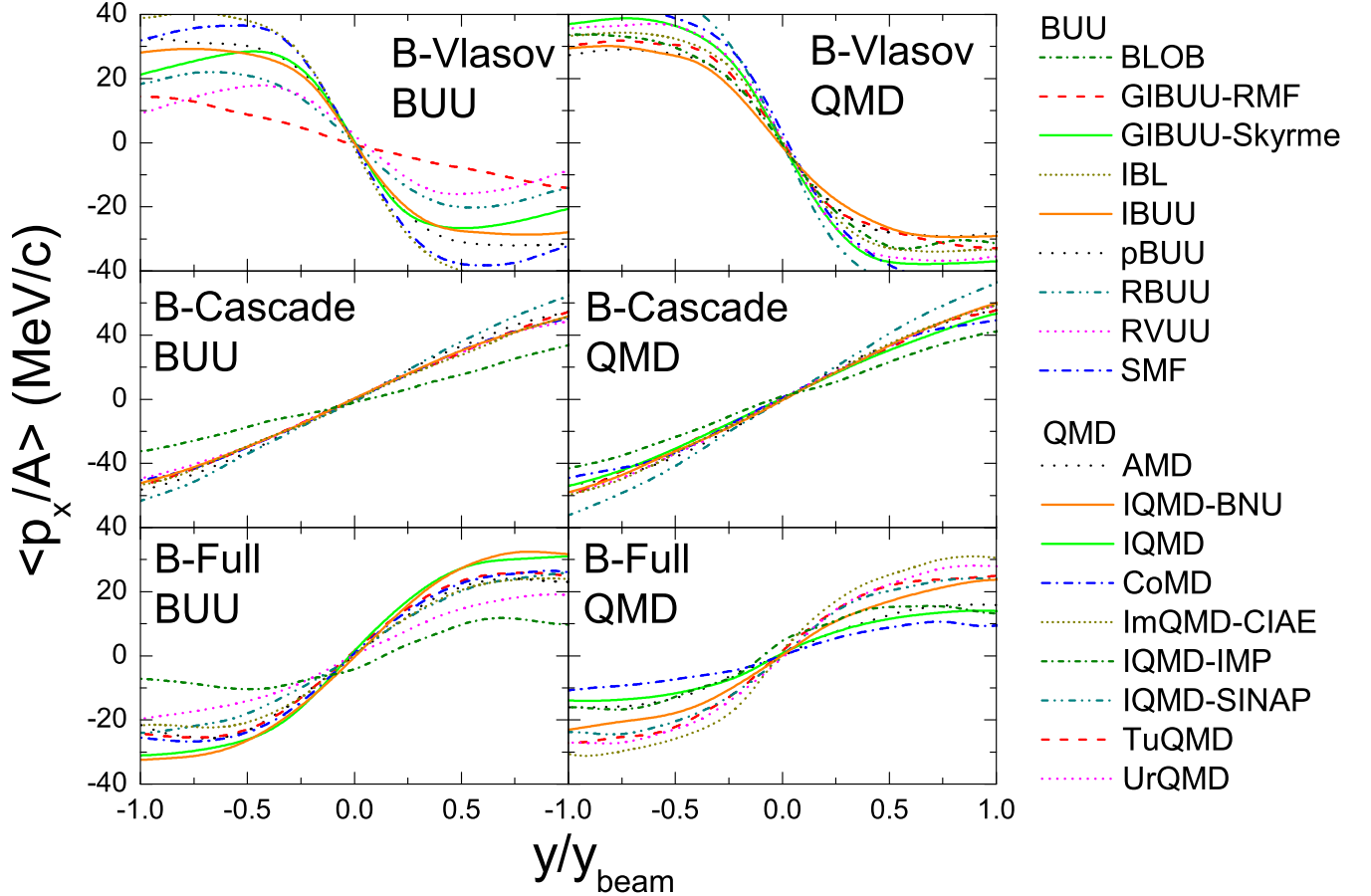}
\caption{(Color online) Transverse flow as a function of reduced
rapidity for the B-Vlasov mode (top panels), the B-Cascade mode
(middle panels), and the B-Full mode (bottom panels) at $\text{b}=7$
fm from BUU-type (left panels) and QMD-type (right panels) models. }
\label{v1_B}
\end{figure}

\begin{figure}[h]
\includegraphics[scale=1.4]{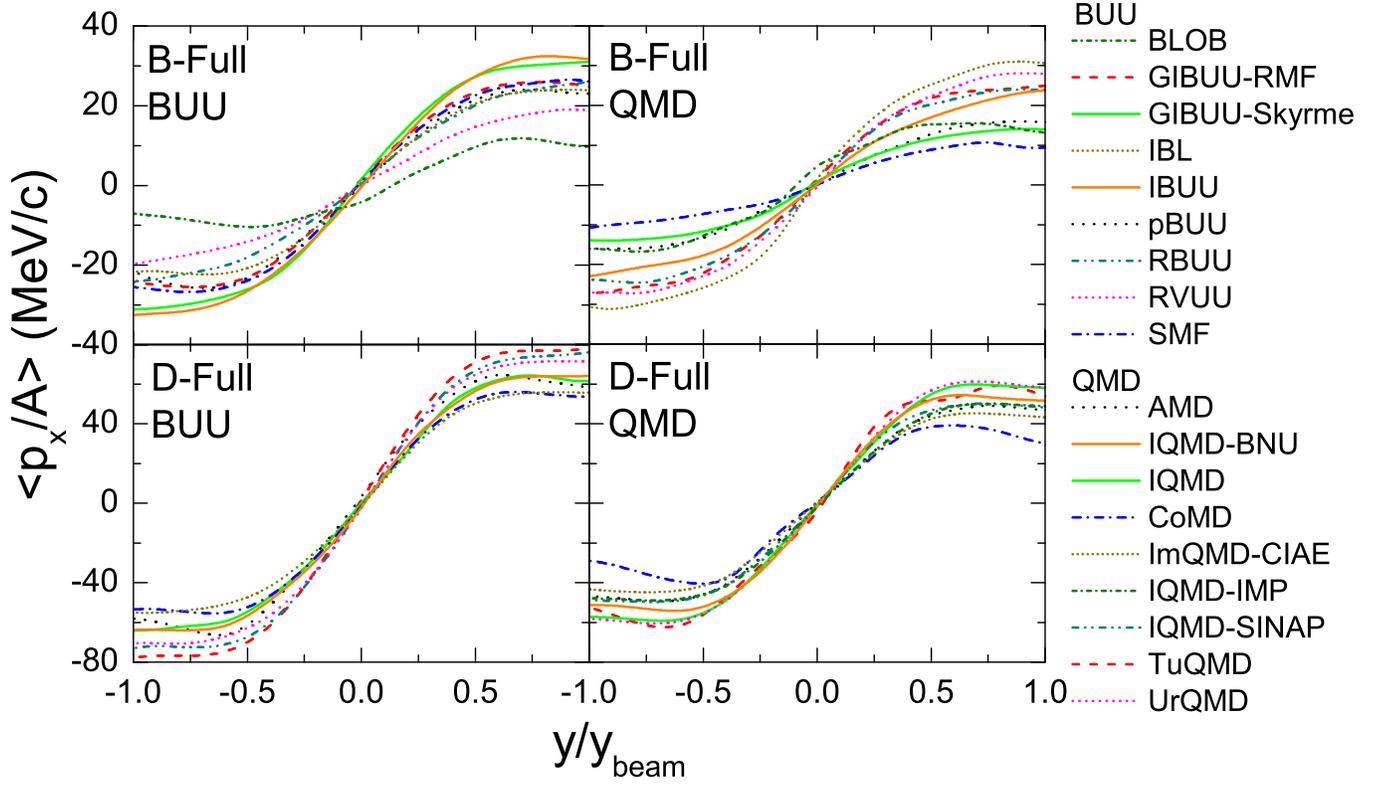}
\caption{(Color online) Transverse flow as a function of reduced
rapidity for the B-Full mode (100 AMeV) and D-Full mode (400 AMeV)
at $\text{b}=7$ fm from BUU-type (left) and QMD-type (right) models.
} \label{v1_BD}
\end{figure}

\begin{figure}[h]
\includegraphics[scale=1.4]{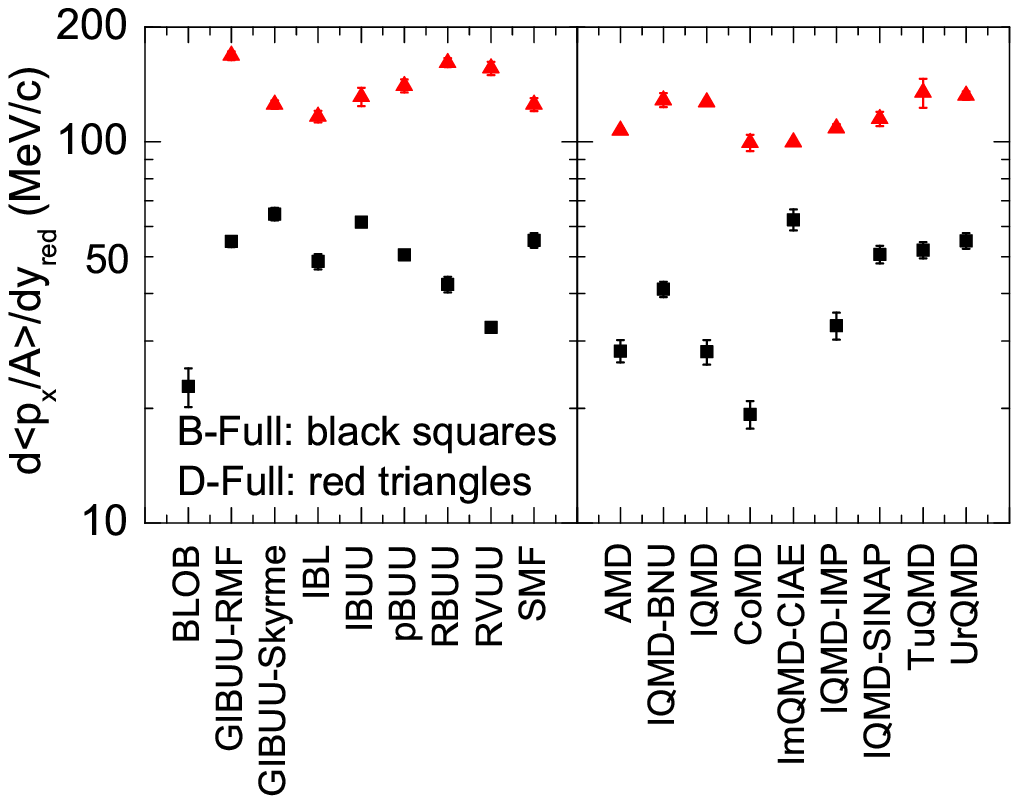}
\caption{Slope parameters of the transverse flow of 9 BUU-type (left) and 9 QMD-type (right) models from the B-Full mode (black squares) and D-Full mode (red triangles). The error bars are the fitting uncertainties. Where they are not seen they are smaller than the symbols. }
\label{flow}
\end{figure}

Given the anisotropy of the evolving density in
Figs.~\ref{denevo_buu} and \ref{denevo_qmd}, one can infer that, in
a collision with finite impact parameter, an anisotropy in the
collective momentum distribution develops. This collective flow is
usually quantified in terms of a Fourier series expansion of the
dependence of the yield  on the azimuthal angle
\begin{equation}
N(\phi;y,p_T)=N_0[1+2\sum_n v_n(y,p_T) \cos(n\phi)].
\end{equation}
The first two coefficients in this expansion ($n=1$ and 2) are
called the directed and elliptic flow, respectively. They are
functions of rapidity $y$ and transverse flow $p_T$ and are important observables in heavy-ion
collisions. They are of particular interest at midrapidity, where
particles and clusters come more directly from the compressed region
and are therefore of interest in the determination of the
high-density EoS. The flow is determined from both the mean field
and NN scatterings, with the relative importance depending on the
incident energy. Here we concentrate on the transverse flow, or
rather on the average in-plane flow $<p_x/A>$, which is closely
related to the directed flow defined as $v_1=< p_x/p_T>$. We do not
show the elliptic flow, normally considered at midrapidity, since
the statistics in our calculations was rather low in the midrapidity
region.

In Fig.~\ref{v1_B} we show the transverse flow per nucleon as a
function of the reduced rapidity $y_{red}=y/y_{beam}$ in the
CM frame for the B-mode calculation from BUU (left) and QMD (right)
codes. The results for Vlasov, Cascade, and Full
modes are shown in the top, middle, and bottom rows, respectively. In
Fig.~\ref{v1_BD} we compare results from the full calculation for the B- and
D-modes (100 and 400 AMeV). Because of the above arguments, the slope
of the transverse flow at midrapidity, often simply called the
¡±flow¡±, is of particular interest. The values of the flow for the
full calculations from a linear fit in the range $|y_{red}| < 0.38$
are plotted in Fig.~\ref{flow}. The error bars shown are the fitting
uncertainties without taking into account the statistical error of
the individual data points. The rapidity range is chosen to reflect
the linear slope region around midrapidity since results from most
codes exhibit an S-shaped curve. Due to symmetry, the transverse flow in Figs.~\ref{v1_B} and \ref{v1_BD} should be zero
for $y = 0$, as is seen in most models. For BLOB the failure to show this
is related to the effectively lower statistics than for other BUU
models, as discussed in Sec.~\ref{fluctuations}, and correspondingly a larger
statistical error.

The transverse flow results from a competition between the mean
field and the scattering terms. Above the Coulomb barrier the mean
field is attractive for some range of energies, which leads to
negative deflection angles, and thus to a negative slope at
midrapidity for the transverse flow, as seen for the B-Vlasov
calculations in Fig.~\ref{v1_B}. The NN scatterings act repulsively
and lead to a positive slope as seen for the B-Cascade mode. The
energy where the two effects just cancel each other is called the
balance energy. We see, in the B-Full mode, that at 100 AMeV we are
somewhat above the balance energy. The slope is positive, but
smaller than that in the Cascade mode (note the change of scale).
One should note that different densities and density profiles are
reached in Cascade, Vlasov, and Full modes, and thus
quantitatively the effects are not directly comparable. At 400 AMeV
the slopes in the full calculation (Fig.~\ref{v1_BD}) are much
larger. At this energy also the mean field acts repulsively (not
illustrated). From these considerations one should expect a
correlation between the flow and the rapidity distribution
(Fig.~\ref{rap_B}), both in B-Vlasov and B-Cascade modes. For the
Full mode this is less clear because of different effects competing.
Indeed, one can observe such a correlation in some cases. Thus, in
B-Vlasov mode RVUU and RBUU show a weak flow and a strong stopping,
while the opposite is true for IBUU. In B-Cascade mode, however,
IQMD-SINAP yields a large flow and a strong stopping, while the
opposite is true for IQMD-IMP. But in other cases the correlations
are not so clear. With increasing incident energy there is a
stronger flow but a weaker stopping, because the competition of the
contributions from the mean field and the NN scatterings to the flow
changes.

For the B-Full mode the spread of the flow is considerable, as seen
in Fig.~\ref{flow}, for both BUU and QMD models, which scatter about
a common value of around 50 MeV/c. The low value observed in the
BLOB case is related to the fact that this model includes
fluctuations in the treatment of the collision integral, leading to
a stronger fragmentation. At the higher incident energy the results
for the transverse flow are visually closer.  By assuming a systematic error of $3\%$ for the transverse flow results besides the fitting error, we obtain the mean
flow of $51 \pm 11$ MeV/c at 100 AMeV and $143 \pm 19$ MeV/c at 400
AMeV for the BUU codes and $45 \pm 13$ MeV/c at 100 AMeV and $116 \pm 12$ MeV/c
at 400 AMeV for the QMD codes. The uncertainties reflect the
standard deviations, which are similar in magnitude but relatively increase by
about a factor of two, by comparing those at 400 AMeV with those at 100 AMeV. The
results for BUU and QMD codes overlap within their error bars, but it seems that QMD models give a systematically smaller flow, which at 400 AMeV amounts to about $15\%$. One
may conclude that flow is a robust observable with uncertainties
from the simulations of around 13$\%$ at 400 AMeV. At the lower
energy of 100 AMeV the uncertainties increase to about $30\%$.

\section{Discussion}
\label{discussion}

Transport theories, in particular the BUU and QMD approaches, have
been widely used in extracting physics information from heavy-ion
collisions. However, because of the complexity of these theories,
the corresponding simulations involve many choices and strategies.
In this paper we have studied the robustness of the predictions from
different transport simulation codes under controlled conditions of
identical physics input and with as close as possible initial
conditions, in order to obtain an estimate of the spread of the
results.

In this comparison we find a considerable spread of results from
different codes, which is larger at the lower incident energy. Since the
¡±philosophies¡± of BUU and QMD approaches are different, e.g.,
fluctuations and correlations are treated differently, one cannot
expect to completely eliminate the difference between the two
approaches, but it is reassuring to see that similar results are
obtained. While we do not attempt to validate different codes,
we try to understand ¡°outliers¡± with the help of the code authors
where possible. The range of results found in this comparison gives
for the first time an estimate of a benchmark for transport codes in
these energy regions. One goal of the project is to improve these
benchmarks, by investigating what causes the diverse results, and to
identify the best strategies and methods to simulate the transport
equations. In the absence of complete agreement, the aim of the
project is to bring about theoretical uncertainties of less than
10$\%$, which are typically achieved in experiments, so that
comparisons of transport results with experimental data become more
meaningful and robust.

 One should consider the spread of the predictions, such as those shown in Fig.~\ref{flow} and earlier figures, as a kind of systematic theoretical error of transport simulations, i.e., if another code were used to interpret the same experimental data, a different conclusion within this systematic uncertainty should be expected. Agreement of a simulation with an experimental observable alone may not serve as validation that the extracted physical parameters are reliable, because the variation of physical parameters could be compensated by strategies in the simulation. Of course, such a compensation is less likely, if a code is able to describe with the same physical input different observables at different energies or different impact parameters.

Over the years, a large amount of the experimental data has been
obtained in the extended Fermi energy domain from about 35 to
 150 AMeV, including data on isotope
yields~\cite{Tsa01,Cha14,Mar12,Wue09,Wad12,Qin12} and on isospin
diffusion~\cite{Tsa04}, which at present have been most widely used
for the determination of the symmetry energy at subsaturation
densities. Our results seem to indicate that the robustness of the
predictions may be reduced in this energy region, which is
particularly sensitive because of competing effects of the mean
field and the NN collisions on the observables and because of the
importance of cluster formation. This is consistent with different
conclusions reached in the transport simulation
analyses~\cite{Che05,Fam06,Riz08,Tsa09,Cou14,Kon15}. Understanding
and improving the predictions here is particularly important and
should be undertaken in the future.

\section{Conclusion and Outlook}
\label{summary}

The goal of this code comparison project is to determine and
ultimately reduce the model uncertainties, in order to extract
model-independent information on nuclear interactions from heavy-ion
experiments. Based on 9 BUU-type models and 9 QMD-type models, we
have compared the initialization, stability of the initial
configuration, the number of attempted and successful collisions, and
the effects of the mean field and collision terms, on the rapidity
distribution and anisotropic collective flows in Au+Au collisions at
beam energies of 100 and 400 AMeV. Although there is still
considerable model dependence that needs to be further understood, we have
learned some useful lessons from the comparison. We have found that
the results from BUU and QMD approaches are essentially similar for
the quantities compared here. The differences in the collision
strategies are less important at higher incident energies as a
result of weaker effects from initialization and Pauli blocking, and
consequently the robustness of the predictions is higher. For the
flow observables, we find uncertainties from code dependence of
about 30$\%$ at 100 AMeV and 13$\%$ at 400 AMeV.

The divergence in the results of different codes very likely
originates from several sources: Firstly, the initialization of the
colliding nuclei is found to be rather different despite attempts to
use a prescribed initial density profile. This seems to lead to
differences in the evolution. Here an initialization based on
approximate ground states consistent with the employed mean fields
should improve the comparison. Secondly, from simulations with
mean-field potentials only and with nucleon-nucleon scatterings
only at 100 AMeV, the results for the attempted collisions and for
the Pauli blocking factors show considerable differences in
different codes. Some of these differences are due to different
physically motivated strategies, which were not prescribed in this
comparison, but left as in the normal usage of the code. This part
of the transport simulations should be critically assessed in future
comparisons.

In heavy-ion collisions many effects are closely intermingled: the
density evolution, the collision probabilities and their Pauli
blocking, the collective flows, and the fragmentation and
clusterization. To understand these different factors better, it is
advantageous to test them separately as much as possible. Therefore
as a follow-up to the present work, plans are being made to compare
simulations from different codes for a system of infinite nuclear
matter, i.e., a calculation in a box with periodic boundary
conditions. Here the initialization should not be problematic, the
overall density is a constant, and the energy conservation should be
strictly obeyed. For many quantities, such as the collision rates,
exact analytic limits are available, and thus one may test the
thermodynamic consistency of the codes, and disentangle better the
effects of the mean field and those of the nucleon-nucleon scatterings.
The results from such comparisons could establish important
benchmarks in understanding transport simulations of heavy-ion
collisions. Another future direction will be to compare more
complicated, but in practice very important aspects of heavy-ion
reactions, such as momentum and isospin dependence of the mean
field, isospin transport, clusterization and fragmentation, and
particle production.

\begin{acknowledgments}
JX acknowledges support from the Major State Basic Research
Development Program (973 Program) of China under Contract No.
2015CB856904 and No. 2014CB845401, the National Natural Science
Foundation of China under Grant No. 11475243 and No. 11421505, the
"100-Talent Plan" of Shanghai Institute of Applied Physics under
Grant No. Y290061011 and No. Y526011011 from the Chinese Academy of
Sciences, the Shanghai Key Laboratory of Particle Physics and
Cosmology under Grant No. 15DZ2272100, and the "Shanghai Pujiang
Program" under Grant No. 13PJ1410600. LWC acknowledges the Major
State Basic Research Development Program (973 Program) in China
under Contract No. 2013CB834405 and No. 2015CB856904, the National
Natural Science Foundation of China under Grant No. 11275125 and No.
11135011, the "Shu Guang" project supported by Shanghai Municipal
Education Commission and Shanghai Education Development Foundation,
the Program for Professor of Special Appointment (Eastern Scholar)
at Shanghai Institutions of Higher Learning, and the Science and
Technology Commission of Shanghai Municipality (11DZ2260700).  MBT
acknowledges support from the USA National Science Foundation Grants
No. PHY-1102511 and travel support from CUSTIPEN (China-US Theory
Institute for Physics with Exotic Nuclei) under the US Department of
Energy Grant No. DE-FG02-13ER42025. HHW acknowledges support from
the DFG Cluster of Excellence {\it Origin and Structure of the
Universe}, Germany. YXZ acknowledges the supports from the National
Natural Science Foundation of China under Grant No. 11475262 and the
National Key Basic Research Development Program of China under Grant
No. 2013CB834404. KK and YK acknowledge the supports from the Rare
Isotope Science Project of Institute for Basic Science funded by the
Ministry of Science, ICT and Future Planning and the National
Research Foundation of Korea (2013M7A1A1075764). CMK acknowledges
support by the Welch Foundation under Grant No. A-1358. BAL is
supported in part by the U.S. National Science Foundation under
Grant No. PHY-1068022, the U.S. Department of Energy¡¯s Office of
Science under Award No. DE-SC0013702, and the National Natural
Science Foundation of China under Grant No. 11320101004. AO
acknowledges support from JSPS KAKENHI Grant No. 24105008. PD
acknowledges support from the USA National Science Foundation Grant
No. PHY-1510971. YJW and QFL acknowledge support by the National
Natural Science Foundation of China under Grant Nos. 11505057,
11547312, and 11375062. GQZ acknowledges support by the National
Natural Science Foundation of China under Grant No. 11205230. The
authors would like to thank all the participants of the Transport
2014 workshop for their enthusiasm, support, and encouragement for
the code comparison project. In particular, we would like to thank
Zhigang Xiao, Yvonne Leifels, and Li Ou for their great help in
leading stimulating discussion sessions during the workshop, and Joe
Natowitz for careful reading of the manuscript and providing
valuable comments. The writing committee (consisting mainly of the
first 5 authors) would like to acknowledge the support and the
excellent hospitality for their writing sessions at the Jagiellonian
University, Cracow, Poland, and at Shanghai Institute of Applied
Physics, Shanghai, China.

\end{acknowledgments}


\begin{thebibliography}{99}
\bibitem{GXNU-IQMD}  The results from ImQMD-GXNU from Ning Wang and Jun-Long Tian are not included in the comparison as this code is developed mainly for very low-energy reactions ($<50$ AMeV).
\end{thebibliography}
\end{document}